\pdfoutput=1

\documentclass[12pt]{article}

\usepackage{amsfonts}
\usepackage{amsmath}
\usepackage{amssymb}
\usepackage{bigints}  
\usepackage{booktabs}
\usepackage[nosort]{cite}
\usepackage{color}
\usepackage{dsfont}
\usepackage{float}
\usepackage{framed}
\usepackage{graphicx}
\usepackage{indentfirst}
\usepackage{mathrsfs}
\usepackage{multirow}
\usepackage{pdflscape}
\usepackage{setspace}
\usepackage{subdepth}
\usepackage{subfig}
\usepackage{titlesec}
\usepackage[dotinlabels]{titletoc}
\usepackage{wrapfig}
\usepackage[all]{xy}
\usepackage{young}
\usepackage[vcentermath]{youngtab}
\usepackage{relsize}
\usepackage{stackengine}
\usepackage{datetime}

\usepackage{hyperref}
\usepackage{physics}
\usepackage[dvipsnames]{xcolor}

\usepackage{mathtools}

\numberwithin{equation}{section}

\usepackage{verbatim}

\newcommand{\be}{\begin{equation}}
\newcommand{\ee}{\end{equation}}

\newcommand{\bml}{\begin{multline}}
\newcommand{\emll}{\end{multline}}

\newcommand{\nn}{\nonumber}

\def\({\left(} \def\){\right)}
\def\[{\left[} \def\]{\right]}

\def\Re{\text{Re}}
\def\Im{\text{Im}}

\def\sgn{\text{sgn}}

\def\al{\alpha}

\def\mI{\mathcal{I}}

\def\eps{\epsilon}

\def\v{\vec}

\def\g{\gamma}

\def\lam{\lambda}

\def\d{\partial}

\def\o{\omega}

\newcommand{\la}{\langle}
\newcommand{\ra}{\rangle}

\newcommand{\bea}{\begin{eqnarray}}
\newcommand{\eea}{\end{eqnarray}}

\usepackage[left=2cm,right=2cm,top=2cm,bottom=2cm]{geometry}
\titleformat{\section}{\large\bfseries}{\thesection.}{4pt}{}
\titlespacing{\section}{0pt}{22pt}{6pt}

\titleformat{\subsection}{\normalfont\bfseries}{\thesubsection.}{4pt}{}
\titlespacing{\subsection}{0pt}{18pt}{6pt}

\titleformat{\subsubsection}{\normalfont\itshape}{\thesubsubsection.}{4pt}{}
\titlespacing{\subsubsection}{0pt}{16pt}{6pt}

\def\ie{\begin{equation}\begin{aligned}}
\def\fe{\end{aligned}\end{equation}}

\def\tilde{\widetilde}
\def\t{\tilde}

\def\d{\partial}

\def\1{{\mathds 1}}
\def\Re{\mathop{\rm Re}}
\def\Im{\mathop{\rm Im}}

\def\mN{\mathcal{N}}

\def\mL{\mathcal{L}}

\def\o{\omega}
\def\v{\vec }

\DeclareFontShape{OT1}{cmr}{mx}{n}%
    {<->cmr10}{}
\newcommand{\mytitlefont}{\fontseries{mx}\selectfont}
\DeclareMathAlphabet{\titlemath}{OT1}{cmr}{mx}{n}

\newcommand{\bi}{\begin{itemize}}
\newcommand{\ei}{\end{itemize}}

\newcommand{\sss}{\subsubsection}

\usepackage{bm}

\usepackage{tikz}
\tikzset{every picture/.style={line width=0.75pt}} 
\begin{document}

\begin{titlepage} 
\begin{center}
~\\[1cm]

{\fontsize{20pt}{0pt} \mytitlefont Weak and strong turbulence in self-focusing and defocusing media}\\[10pt]

~\\[0.2cm]

{\fontsize{14pt}{0pt}Vladimir Rosenhaus{\small $^{1}$} and Gregory Falkovich{\small $^{2}$}}
~\\[.3cm]

\it{$^1$ Initiative for the Theoretical Sciences}\\ \it{ The Graduate Center, CUNY}\\ \it{
 365 Fifth Ave, New York, NY 10016, USA}\\[.5cm]
 
 \it{$^2$Weizmann Institute of Science}, \it{Rehovot 76100 Israel}
\\[25pt]\end{center}
%\today

\noindent

While the focusing and defocusing Nonlinear Schr\"odinger Equations have similar behavior in the weak turbulence regime, they must differ dramatically in the strong turbulence regime. 
Here, we show that this difference is already present at  next-to-leading order in the nonlinearity in the weak turbulence regime: The one-loop correction to the interaction vertex suppresses repulsion (like screening in electrodynamics), leading to a  steeper spectrum in the defocusing case. In contrast,  attraction enhancement (like antiscreening in chromodynamics) makes the spectrum less steep in the focusing case.

 To describe strong turbulence, we consider a vector model in the limit of a large number of components. 
A large-$N$ kinetic equation, valid at all scales, can be derived analytically. It has an inverse-cascade solution whose two asymptotics, at high and low wavenumbers, describe weak and strong turbulence, respectively.  We find two forms of universality  in the strong turbulence spectrum: in focusing media it is independent of the flux magnitude, while in  defocusing media it is independent of the bare coupling constant,  with the largest scale appearing instead.
\\

\vfill 
\today
\end{titlepage}
\tableofcontents
~\\[-20pt]
\vspace{.2cm}
\section{Introduction}

The Nonlinear Schr\"odinger Equation (NSE) is arguably the most universal model in physics, describing spectrally narrow wave packets and long-wavelength limits of systems of waves and particles, see e.g. \cite{OT,Naz,ZLF,exp1,exp2,exp3,exp4,Nowak:2011sk,Pom,B15,Naz23,Pre,PT,Gaz19}:
\be 
i \Psi_t=\frac{\delta{\cal H}}{\delta\Psi^*}~,\  \ \ \ \ \ {\cal H}=\int d{\bf r}\left[|\nabla\Psi|^2+\lambda|\Psi|^4\right]\,.\label{NSE}
\ee
The first term in ${\cal H}$ can be interpreted as the kinetic energy, while the second term describes an  interaction that is local in ${\bf r}$-space: either repulsion for $\lambda>0$ or attraction for $\lambda<0$. The latter describes, in particular, self-focusing of light in nonlinear media. The conserved quantities are the energy ${\cal H}$ and the wave action ${\cal  N}=\int d{\bf r}|\Psi|^2/V$, where $V$ is the  volume. 

The ratio of the interaction energy to the kinetic energy defines the dimensionless nonlinearity parameter, $\epsilon_k=\lambda n_kk^{d-2}$,  which is expressed here via the occupation number $n_k=\langle |\Psi_k|^2\rangle$ of mode $k$. A necessary, but insufficient, condition for weak nonlinearity is $\epsilon_k\ll1$. In this case, the main part of the energy spectral density is $k^2n_k$. If we pump both energy and action at a finite scale, dissipation at large $k$ absorbs a finite amount of energy and very little action. Therefore, weak turbulence has a direct cascade of energy and an inverse cascade of wave action. This argument might be extendable to strong turbulence with $\lambda>0$, where the interaction adds to the energy. In contrast, for $\lambda<0$ strong turbulence can burn finite amounts of both energy and action at large $k$.

An inverse cascade carries a constant spectral flux of action, $Q$, which can be estimated as the spectral density, $ n_kk^d$, divided by the weakly nonlinear interaction time,  $\tau_k\simeq 1/\omega_k\epsilon_k^2$, where $\o_p =p^2$ is the dispersion relation. This gives  the Kolmogorov-Zakharov spectrum (KZ state),
\be  
 n_k=k^{-d+2/3}(Q/\lambda^2)^{1/3}\,,\label{KZ}
\ee
which corresponds to $\epsilon_k\propto \lam k^{-4/3}$ --
an  inverse cascade inevitably transforms  turbulence from weak to strong for small enough $k$. For weak turbulence, one can systematically compute  the spectrum in powers of $\eps_k$ \cite{RS1, RS2, RSSS}. In Sec.~\ref{sec2} we study the first order correction, showing that it leads to a spectrum that is steeper than (\ref{KZ}) for positive $\lam$, and shallower than (\ref{KZ}) for negative $\lam$. 
Going into a region where $\eps_k$ is of order one or larger, one expects  on physical grounds that strong turbulence depends crucially on the sign of $\lambda$. For  positive $\lam$, an inverse cascade tends to create a stable and growing condensate that is as large as the box, and a steady state in a finite box needs some large-scale mechanism for absorbing the wave action. On the other hand, for negative $\lam$ attraction leads to instabilities and self-focusing of sufficiently long waves, which carries the action directly from small to large $k$, potentially providing a steady state without large-scale dissipation  \cite{OT}.

An analytic description of strong turbulence is, in general, impossible unless one introduces an additional small parameter. The inverse of the spatial dimension $1/d$ is not such a parameter,  as  $\epsilon_k$ is determined by $n_kk^d$, which is independent of $d$. Instead, in Sec.~\ref{sec3} we consider the generalization of the model to an $N$-component vector $\vec \Psi$. The results of Sec.~\ref{sec2}, on the steepening of the spectrum for positive $\lam$, continue to hold for any $N$. The large-$N$ limit (small $1/N$) allows one to derive a kinetic equation \cite{Nowak:2011sk,B15,Gaz19, RSch}, which admits solutions describing both weak and strong turbulence. The Kolmogorov-Zakharov spectrum (\ref{KZ}) is a large-$k$ asymptotic solution for any sign of $\lambda$. For low $k$ (strong turbulence) in the defocusing case, we argue that the spectrum is given by (\ref{ss1}) and (\ref{ss3}) for freely-decaying and forced turbulence, respectively. Both spectra are independent of $\lambda$. Notably, the spectrum differs from a guess for strong turbulence that is commonly discussed in the literature, $ n_k\simeq k^{-d}(Q/\lambda)^{1/2}$, which is argued for by assuming that the cascade is qualitatively similar to weak turbulence but with the weak-turbulence time replaced by the nonlinear time $1/\omega_k\epsilon_k=1/\lambda_kn_kk^d$. The  strong turbulence regime of the focusing case requires further study of the large-$N$ kinetic equation. We argue that the critical balance scenario  \cite{Phil,Goldreich,Newell,NS}, which corresponds to the universal (flux-independent) spectrum $ n_k\simeq k^{2-d}/\lambda$ and  results from setting $\epsilon_k$ to be a constant, is more likely to occur in two dimensions  than in three dimensions.

\section{Weak Turbulence} \label{sec2}
In this section we consider the limit $\lambda\to0$ and derive the kinetic equation, which describes the evolution of occupation numbers, up to  order $\lam^3$. The equations of motion relate the rate of change of the occupation number $n_1$ of mode $p_1$ to the imaginary part of the fourth moment, where $\Psi_i \equiv \Psi(p_i)$ denotes the Fourier transform of $\Psi(x)$, and $\v p_{12;34} \equiv p_1{+}p_2{-}p_3{-}p_4$:
\be \label{eomKE}
\frac{\d n_1}{\d t} \equiv I_{p_1}=4\lam\int d\v p_2 d\v p_3 d\v p_4\,\delta(\v p_{12;34})\Im \la \Psi_1^*\Psi_2^*\Psi_3\Psi_4\ra\,.~
\ee
The fourth moment is evaluated perturbatively in $\lam$, in a state that is taken to be near Gaussian, with a variance that is self-consistently taken to be $n_k$. Specializing the general results for these higher-order terms  \cite{RS1,RS2,RSSS} to the nonlinear Schr\"odinger equation in three dimensions ($d=3$) gives, to order $\lam^3$, see Appendix~\ref{apA}, 
\be
\frac{\d n_1}{\d t}=16\pi \lam^2 \Re \int d\v p_2 d\v p_3 d\v p_4\,  n_1 n_2 n_3 n_4\Big( \frac{1}{n_1} {+} \frac{1}{n_2}{-}\frac{1}{n_3} {-} \frac{1}{n_4} \Big)\\
\[1 + 2\mL_+ + 8\mL_-\] \delta(\o_{p_1p_2; p_3 p_4})\delta(\v p_{12;34}) + \ldots~,\label{30}
\ee
where $\o_p = p^2$, $\o_{p_1p_2; p_3 p_4}\equiv \o_{p_1}{+} \o_{p_2}{-}\o_{p_3}{-}\o_{p_4}$, and 
\bea \nn
\text{Re } \mL_+ &=& \frac{4\pi \lam}{|\v p_1{+}\v p_2|} \int_0^{\infty} dq\, q\, n_q \log\Big|\frac{q^2 - q |\v p_1{+}\v p_2| +\v p_1{\cdot} \v p_2}{q^2 + q |\v p_1{+}\v p_2|+\v p_1{\cdot} \v p_2}\Big| \\
\text{Re } \mL_- &=& \frac{2\pi \lam}{|\v p_2{-}\v p_4|} \int_0^{\infty} dq\, q\, n_q \log\Big| \frac{q^2 (\v p_2 {-}\v p_4)^2- q |\v p_2{-}\v p_4|^3 -\v p_2{\cdot}(\v p_2{-}\v p_4)\v p_4{\cdot}(\v p_2{-}\v p_4)}{q^2 (\v p_2 {-}\v p_4)^2+ q |\v p_2{-}\v p_4|^3 -\v p_2{\cdot}(\v p_2{-}\v p_4)\v p_4{\cdot}(\v p_2{-}\v p_4)}\Big|~,   \label{loop}
\eea
where  we have written the expression for $\mL_{\pm}$  after performing an internal angular integral,  assuming an isotropic $n_q$. We note that the frequency renormalization expected at this order in $\lam$, which here is just a shift by a constant, $\omega_k\rightarrow\o_k+\lambda \mN$, doesn't affect the kinetic equation, which only involves differences of frequencies. 
 The term in square brackets in (\ref{30}) contains three terms: the $1$, which by itself gives the ``standard'' wave kinetic equation, and the $\mL_{\pm}$ terms, which are suppressed by an additional power of $\lam$ and scale as $\eps_k$. The standard kinetic equation admits a steady-state weakly turbulent  solution  (\ref{KZ}), and we would like to understand how the small corrections $\mL_{\pm}$ alter this solution.  
 
Schematically, one expects that the stationary turbulent solution will have a power series expansion in $\eps_k$, in proximity to the KZ state (\ref{KZ}), 
\be
n_k \sim k^{-d + 2/3} \(1 + \# \lam k^{-4/3} + \ldots \)~,
\ee
although we will see later that the situation is more subtle. Ideally, we would like to establish the number in front of the $\lam$. However, our goal now is more modest: we simply want to establish if it is positive or negative. 

To the order $\lam^3$ to which we are working, we may replace the $n_k$ in $\mL_{\pm}$ by the KZ state (\ref{KZ}) and perform the integral over the magnitude of the momentum, giving, 
 \bea 
\text{Re }\mL_+ &=& -\frac{4\pi^2 2^{\frac{1}{3}}\sqrt{3} \lam}{|\v p_1{+}\v p_2|} \( \t p_+^{\, -1/3}+ \t p_-|\t p_-|^{-4/3}\),~\ \ \ \ \ \ \ \ \tilde p_\pm\equiv | \v p_1{+} \v p_2|\pm  | \v p_1{-} \v p_2|\label{plus}\\
\text{Re }\mL_- &=&- \frac{\pi^2 2\sqrt{3} \lam}{|\v p_2{-}\v p_4|^{2/3}}\( \frac{ p_4^2{ -} \v p_2 {\cdot} \v p_4}{|p_4^2 {-} \v p_2 {\cdot} \v p_4|^{4/3}}+ \frac{ p_2^2{ -} \v p_2 {\cdot} \v p_4}{|p_2^2 {-} \v p_2 {\cdot} \v p_4|^{4/3}}\)~.\label{minus}
 \eea
 We may write the kinetic equation (\ref{30}) in the form 
   \be \label{27}
\frac{\d n_1}{\d t} = 16\pi \text{Re}\, \int d\v p_2 d\v p_3 d\v p_4\,  n_1 n_2 n_3 n_4  \Lambda_{1234}^2 \Big( \frac{1}{n_1} {+} \frac{1}{n_2}{-}\frac{1}{n_3} {-} \frac{1}{n_4} \Big)  \delta(\o_{p_1p_2; p_3 p_4})\delta(\v p_{12;34}) 
\ee
so that it looks like the standard kinetic equation but with an effective (or renormalized) interaction $\Lambda_{1234}^2 = \lam^2 \(1 + 2\mL_+ + 8\mL_-\)$. More precisely, we should symmetrize this equation, to give $\Lambda_{1234}$ the symmetry properties of an interaction, see Appendix~\ref{apA}. 
While our original interaction was simply a constant, the renormalized interaction --  and even the sign of the $\mL_{\pm}$ contributions -- is momentum-dependent. Indeed, the $\tilde p_-$ term in $\mL_+$, as well as  both terms in  $\mL_-$, changes signs upon passing through the resonances $| \v p_i- \v p_j|^2=p_i^2-p_j^2$. By way of interpretation, this seems analogous to the response of a harmonic oscillator  changing sign when the driving frequency passes through a resonance. The role of the driving force is played by the intermediate waves inside the loop,  having momenta $p_5,p_6$ in the notation of Fig.~\ref{tree} in Appendix~\ref{apA}.

Since $n_k$ is assumed to be isotropic, we perform the angular integrals in (\ref{27}), transforming it into, 
 \be \label{B2}
S_d\, \frac{\d n(\o_1)}{\d t}  =16\pi\, \text{Re} \int\!\! d\o_2 d\o_3 d\o_4 \Big( \frac{1}{n_1} {+} \frac{1}{n_2}{-}\frac{1}{n_3} {-} \frac{1}{n_4} \Big) n_1 n_2 n_3 n_4 \delta(\o_{12;34})U(\o_1, \o_2, \o_3, \o_4)~,
 \ee
 where $S_d$ is the $d$-dimensional solid angle and $U$ includes the angular integrals of the effective interaction,
 \be \label{U1234}
U(\o_1, \o_2, \o_3, \o_4)  =  \prod_{i=2}^4 \frac{d p_i}{d\o_i} p_i^{d-1}\,\int d\Omega_1 d\Omega_2 d\Omega_3 d\Omega_4 \Lambda_{1234}^2  \delta(\v p_{12;34})~,
 \ee
 where $d\Omega$ is the integral over the sold angle. It is obvious that the contribution to $U$ coming from the $\lam^2$ piece of $\Lambda_{1234}^2$ is nonnegative,  
 \be
 \int d\Omega_1 d\Omega_2 d\Omega_3 d\Omega_4  \delta(\v p_{12;34}) \geq 0~.
 \ee
 We demonstrate in Appendix~\ref{apA4} that the contributions of $\mL_{\pm}$ always have a sign that is opposite to that of $\lam$,
\be \label{sgnA}
\sgn \int d\Omega_1 d\Omega_2 d\Omega_3 d\Omega_4\, \text{Re }\mL_{\pm} \delta(\v p_{12;34}) = - \sgn \lam~.
 \ee
 This is true for all $\o_i$ arguments. This result is consistent with the hypothesis put forward in \cite{FR}, that the criterion for  interaction enhancement is $\lambda d^2\omega_k/dk^2<0$. In our case, $\omega_k=k^2$, attraction ($\lambda<0$) is enhanced and repulsion ($\lam>0$) is suppressed by the one-loop corrections.

The flux is defined as
\be
 Q(k) =- \int_0^k d^d p_1 \frac{\d n_1}{\d t}~.\label{Q}
 \ee 
For $\lam>0$,  the contribution to  $U$ from the corrections $\mL_{\pm}$ is negative for all $\o_i$.  
Inserting  (\ref{B2}) into (\ref{Q}), we see that if $U$ is lower, then  $n_k$ must be higher than the KZ spectrum to maintain a constant flux.
Since the corrections increase in magnitude as we go along the cascade to lower $k$,   the spectrum rises more steeply than the KZ spectrum  in the defocusing case. Similarly, interaction enhancement  requires lower occupation numbers and a less steep spectrum  in the focusing case. 

It is instructive to compare the interaction renormalization in turbulence with that  in thermal equilibrium. In the latter case, the imaginary part of the fourth moment (determining the flux) is identically zero at all orders in $\lambda$. The mean interaction energy  is the real part of the fourth moment,
$U=\lambda\sum_{1234}\Re \la\Psi_1\Psi_2\Psi_3^*\Psi_4^*\ra\delta(\v p_{12;34})=2\lambda {\mN}^2 V+O(\lambda^2)$, which has the same sign as $\lambda$.

\section{Strong turbulence} \label{sec3}
For an analytic description of strong turbulence, we consider a vector variant of the nonlinear Schr\"odinger model (\ref{NSE}), 
\be \label{largeNH}
\ {\cal H}=\int d{\bf r}\left[|\nabla \v \Psi|^2+\frac{\lambda}{N} (\v \Psi^* {\cdot} \v \Psi)^2\right]~,
\ee
in which the number of components $N$ of $\v \Psi$ is taken to be large. Large $N$  is a powerful  technique, widely used to make field theories tractable at strong coupling \cite{Coleman, Wilson:1973jj, PhysRevD.10.3235,  Rosenhaus:2018dtp, Klebanov:2018fzb}. In the context of the NSE,  the different components correspond, for instance, to different directions in which a spin may point, see e.g., \cite{Schumacher:2023rkj} for $N=3$. Of course, we cannot a priori know to what extent strong turbulence in the large $N$ NSE is reflective of strong turbulence in the $N=1$ NSE. A natural first step, however, is to verify qualitative agreement in the weak turbulence regime. In Appendix~\ref{apB} we generalize the kinetic equation to order $\lam^3$ discussed in the previous section to the case of any finite $N$, see (\ref{KEnlo}). Since the only change is that the  coefficients of $\mL_{\pm}$ in (\ref{30}) become $N$-dependent positive constants, the results for the $N=1$ case hold for all $N$: in the defocusing and focusing cases the spectrum is steeper or less steep than KZ scaling, respectively. 

The power of large $N$ is that one can find the kinetic equation to all orders in $\lam$,  at leading nontrivial order in $1/N$. This is achieved by summing a geometric series of ``bubble diagrams''. Roughly, each bubble contributes a factor of $\mL_-$ (encountered earlier in the one-loop kinetic equation (\ref{loop})) and the resulting sum produces a kinetic equation resembling the standard (leading order in $\lam$) kinetic equation, but with an effective coupling $\Lambda_{1234}$, see \cite{bergesGasenzerScheppach2010, berges2002,bergesRothkopfSchmidt2008, Walz:2017ffj, Nowak:2011sk,B15, RSch}, 
   \be  \label{largeNKE}
\!\!\frac{\d n_1}{\d t} = \frac{8\pi}{N}\, \text{Re}\! \int d\v p_2 d\v p_3 d\v p_4\,  n_1 n_2 n_3 n_4 | \Lambda_{1234}|^2 \Big( \frac{1}{n_1} {+} \frac{1}{n_2}{-}\frac{1}{n_3} {-} \frac{1}{n_4} \Big)  \delta(\o_{12;34})\delta(\v p_{12;34})\equiv \tilde I(p_1)\,, \ \ \  
\ee
where
\be \label{Lambda1234}
| \Lambda_{1234}|^2 = \frac{\lam^2}{|1 - \mL_-|^2}~.
\ee
The form of $\mL_-$ in general dimension is given in Appendix~\ref{apA}, (\ref{mLm1}). In three dimensions \cite{Walz:2017ffj}, see (\ref{Lminus0}), 
 \be \label{Lminus03d}
 \mL_- =   \frac{2\pi \lam}{p_-} \int_0^{\infty} dq\, q\, n_q \log\Big| \frac{((p_-{-}q)^2{-}q^2)^2 {-}\o_-^2}{((p_-{+}q)^2{-}q^2)^2 {-}\o_-^2}\Big|  - i\frac{2 \pi^2 \lam}{p_-} \int_ \frac{|p_-^2-\o_-|}{2 p_- }^{ \frac{|p_-^2+\o_-|}{2 p_- }} dq\, q\, n_q~,
 \ee
 where $\v p_- = \v p_4 - \v p_2$ and $\o_- = \o_{p_4} - \o_{p_2}$ . The real part of $\mL_-$ was written previously in (\ref{loop}). 
 
  The large-$N$ kinetic equation (\ref{largeNKE}) is a closed integral equation. It should be possible to numerically solve it for both time-dependent and stationary solutions, at all values of $k$. This task is left for future work. Here, we aim to make analytic progress in understanding it. Notice that 
 if $\mL_- \ll 1$ then $\Lambda_{1234} \approx \lam$, placing us in the regime of weak turbulence. The transition from weak to strong turbulence may be expected to occur as $k$ decreases below the $k_*$ at which $\mL_-$ is of order $1$.  The scaling of $\mL_-$ is $\mL_- \sim \lam k^{d-2} n_k$, which, upon substituting the KZ solution (\ref{KZ}), gives $k_*\simeq (\lambda Q)^{1/4}$. 
 
 In what follows, we will focus on three dimensions, occasionally writing $d$ in some equations that generalize to other dimensions. We will comment on the two-dimensional case in Sec.~\ref{sec2d}.  
  An important feature of the solution to the large $N$ kinetic equation will be the appearance of UV and IR cutoffs.
  For convergence analysis, we need the asymptotics of the integrand in $\text{Re }\mL_-$ (\ref{Lminus03d}) in the regions of small and large $q$: 
 \bea \label{ReLsmallq0}
 - \log\Big|\frac{(p_-^2{-}2 p_- q)^2 -\o_-^2}{(p_-^2{+}2 p_- q)^2 -\o_-^2}\Big|  \approx \begin{cases} &  \frac{8 q p_-^3}{p_-^4{ -} \o_-^2}~, \ \ \ q\ll p_-, \o_-\\ 
 & \frac{2 p_-}{q}~, \ \  \ \ \ \ q\gg p_-, \o_-~.
 \end{cases}
 \eea
 If $n_k$ is a power law, $n_k \sim k^{-\g}$, then $\text{Re }\mL_-$ diverges in the IR if $\g> d$. This divergence would have to be explicitly cut off with an IR cutoff. In the UV, there is a divergence if $\g\leq d{-}2$. This divergence would be cutoff by the scale $k_*$, where the transition from strong turbulence to the KZ solution (valid for  $ k > k_*$) occurs. The KZ    exponent, $\g = d{-} 2/3$, ensures UV convergence. We will find that  interactions with long waves (signified by an IR divergence)  are relevant for the defocusing case, while  interactions with short waves are relevant for the focusing case. 

\subsection{Defocusing ($\lam>0$)}    
In the weak coupling analysis in the previous section,  (\ref{sgnA}) demonstrated that the order $\lam^3$ correction to the kinetic equation resulted in a reduction of the (angle-integrated) coupling (\ref{U1234}) if $\lam$ is positive. It is natural to assume that  interaction suppression persists into the regime of  strong turbulence, which is achieved with either large $\lam$ or small $k$. In this regime,  $\mL_-$ is expected to be large (as it is proportional to $\lam$), making it tempting to neglect the $1$ in the denominator of the effective interaction $\Lambda_{1234}$ (\ref{Lambda1234}), 
\be \label{Lamstrong}
|\Lambda_{1234}|^2 \approx \frac{\lam^2}{|\mL_-|^2}~.
\ee
This would indicate that $\Lambda_{1234}$ decreases as one moves along the inverse cascade to lower $k$. 

\subsubsection{Stationary solution}
Let us begin by searching for the steady-state, constant-flux solution of the large $N$ kinetic equation in the strong turbulence regime. On the basis of power counting, and using (\ref{Lamstrong}), we have   $Q\simeq k^dI_k\simeq k^{3d-3\gamma-2}/|\mL_-|^2\simeq k^{d-\gamma+2}$.  Setting the flux to be constant yields  \cite{Nowak:2011sk}:
\be n_q= Qq^{-d-2}\,.\label{ss2}
\ee
Let us show that this solution cannot be correct. Inserting it into $\mL_-$ leads to an IR divergence, as the exponent $d+2$ exceeds $d$, see below (\ref{ReLsmallq0}). This necessitates the introduction of an IR cutoff $k_0$,  thereby violating the above power counting for the scaling of the flux with $k$. Indeed, using (\ref{ReLsmallq0}), we obtain
   \be \label{117}
   \text{Re}\,  \mL_- \approx - \frac{2 \pi \lam}{p_-} \frac{8  p_-^3}{p_-^4 - \o_-^2} \int_{k_0} \frac{dq}{q} q^{3-\g} = \frac{16\pi \lam  p_-^2}{p_-^4 {-} \o_-^2} \frac{k_0^{3-\g}}{3{-}\g} = \frac{4\lam \mN p_-^2}{\o_-^2{-}p_-^4}~, \ \ \ \ \ d=3~,
   \ee
   where $k_0$ is the IR cutoff, and in the last equality we express $k_0$ in terms of $\mN$, the total number of waves. We consider small $p$, where the nonlinearity is large, and where we are seeking the strong turbulence solution. 
The effective interaction $\Lambda_{1234}$ (\ref{Lamstrong}) becomes, 
\be  \label{BigLam}
|\Lambda_{1234}| = \frac{|\o_-^2{-}p_-^4|}{4 \mN p_-^2}~.
\ee
We notice that, while the interaction in the original nonlinear Schr\"odinger equation was independent of wavenumbers, the effective interaction appearing in strong (large $N$) turbulence has a scaling exponent of two, $\Lambda_{1234} \sim p^2$. This allows  us to show explicitly that (\ref{ss2}) is not a stationary solution of the large-$N$ kinetic equation with   (\ref{BigLam}) at $|\mL|\gg1$. For isotropic spectra, the angle-averaged collision integral  can be written in terms of frequencies $x=\omega_2/\omega_k$, $y=\omega_4/\omega_k$, see (3.1.19) in \cite{ZLF}:
\be
 \!\!\tilde I(k)\sim\frac{k^{2d-3\gamma+2}}{\mN^2}\int_0^1\! dx\int_{1-x}^1\!\!dy\, U\bigl[1{+}(x{+}y{-}1)^{\gamma/2}{-}x^{\gamma/2}{-}y^{\gamma/2}\bigr]
 \bigl[1{+}(x{+}y{-}1)^{\gamma'/2}{-}x^{\gamma'/2}{-}y^{\gamma'/2}\bigr]\,,\label{coll}
 \ee
where $U\equiv U(x,y,1{-}x{-}y)$ is a nonnegative function (which comes from an  angular integral of the square of the effective interaction) and $\gamma'=3(\gamma{-}d){+}2{-}2\beta=3(\gamma{-}d){-}2$, where $\beta$ is the scaling exponent of the interaction, and we took $\beta = 2$ as a result of (\ref{BigLam}). It follows from the identity $\sgn \[ 1+ (y_2 {+}y_3{ -} 1)^{z} {-} y_2^{z} {-} y_3^{z}\] =\sgn\,z(z -1)$ that the collision integral is strictly positive for the $\g = d{+}2$ in (\ref{ss2}). In $d=2$, where  $\gamma=\gamma'=4$, this is clear even without the identity. A strictly positive collision integral is inconsistent with a stationary solution, which requires the integral to vanish. 
 
Finding the correct stationary solution is straightforward, once we accept that it involves the IR cutoff $k_0$. We simply use the effective interaction (\ref{BigLam}) $\Lambda_{1234} \sim k^2$ and apply the standard wave turbulence argument for the scaling of the KZ solution. This gives an exponent $\gamma=d+2/3$. Indeed,  $\gamma'=0$ turns \eqref{coll} into zero and gives  $\gamma=d+2/3$. Thus, in the strong turbulence regime of the defocusing case, the stationary solution is 
  \be n_k\simeq Q^{1/3}{\cal  N}^{2/3} k^{-d-2/3}\simeq Qk_0^{-4/3} k^{-d-2/3}\,.\label{ss3}\ee
  
This formula is the main result of this section. 
It gives an $n_k$ that is larger than the KZ spectrum $n_k\simeq (Q/\lambda^2)^{1/3}k^{-d+2/3}$ by a factor of $(k_*^2/k_0k)^{4/3}$. The enhancement is  both due to $k_0$ appearing with a negative exponent and as a result of the scaling exponent $d{+}2/3$ being larger than the KZ exponent $d{-}2/3$. This result is consistent with the weak turbulence analysis in Sec.~\ref{sec2}, combined with Appendix~\ref{apB}, which shows   that as $k$ decreases the spectrum $n_k$ becomes steeper than KZ, for any value of $N$ (the number of components of the field). Now, we see that (for large $N$ at least) as one moves into the deep IR, the spectrum must saturate to a power law with a larger exponent. 
The spectrum (\ref{ss3}) is also steeper than a common guess for what strong turbulence scaling should be, $n_k\simeq k^{-d}(Q/\lambda)^{1/2}$. We may interpret this guess as being incorrect because it doesn't account for the fact  that the nonlinear interaction rate is parametrically slower than $\lambda n_kk^d$ due to interaction suppression. The ultimate interaction vertex suppression  (\ref{BigLam}) can be guessed on the basis of $k^2/\mN$  being the only combination having the dimensionality of $\lambda$. 

 The scaling   $\gamma{=}d{+}2/3$ appeared earlier in \cite{Gaz19} (Eq.~134) in the context of  self-similar regimes of propagation. We stress  that, while normally the stationary constant-flux cascade is established by a front propagating from the UV to the IR, here, due to the presence of the IR cutoff $k_0$, this cannot be the case. The constant-flux solution  must  instead arise after a  second stage of propagation, triggered at the IR sink and propagating backwards into the UV. 
 The presence of $k_0$ in $n_k$ means that the cascade depends on its destination -- a surprising result, and perhaps the first such example in turbulence. 
 
 The interaction vertex non-locality that gives (\ref{ss3})  changes the validity condition for the weak-turbulence spectrum \eqref{KZ}. 
It was pointed out in  \cite{FR} that the effective nonlinearity parameter is the loop integral $\mL$ rather than the  dimensional estimate $\epsilon_k$ \cite{ZLF,Naz}. Therefore, the validity condition for  \eqref{KZ}  is not $\epsilon_k= (k_*/ k)^{4/3}\ll1$,  but rather $\mL \ll1$. Since the loop integral is over all wavenumbers,  a steeper strong--turbulence spectrum adds a diverging ($k_0$-dependent) contribution. As a result, $\mL\simeq  (k_*^2/k_0 k)^{-2/3}$, which is of order unity  already at $k_1\simeq k_*^2/k_0$. Therefore,  weak turbulence ends not at $k_*$, but at the much larger $k_1$, which depends on $k_0$ for a long cascade with $k_0\ll k_*$.  Of course, if the cascade is short and ends at $k_0>k_*$, then the cascade is entirely within the weak--turbulence domain. As shown in the preceding section, the loop integral converges in this case, so that $\mL\simeq \epsilon_k$.  In the defocusing case, it seems natural that a long enough inverse cascade produces long pieces of quasi-condensate with compact vortices and shocks \cite{Pre}, which can alter the interaction between short waves.

  \subsubsection{Self-similar solutions} \label{selfMain}
    One expects that either at very weak coupling or at very strong coupling, and far away from the forcing and dissipation scales, the kinetic equation should have solutions that are scale invariant, taking the form, 
\be 
n_k = t^a f(k t^b)~.
\ee
Such self-similarity is expected to appear in the long-time limit and requires large boxes.  One can insert this form of the solution into the large $N$ kinetic equation and solve it numerically  \cite{Nowak:2011sk,B15}; the structure of the solution can be readily understood.\\[-5pt]

\textit{Freely-decaying turbulence: }
 As reviewed in Appendix~\ref{apD}, for the standard kinetic equation with a dispersion relation $\o_k = k^2$ and an interaction that scales as $\Lambda_{1234} \sim k^{\beta}$, the exponents $a$ and $b$ for freely decaying turbulence (without external pumping and dissipation) are given by $a = \frac{d}{2\beta - 2}$ and $b = \frac{1}{2\beta - 2}$. Taking $f(\xi) \sim k^{-\g}$ gives  a moving front solution,
  $n_k \sim t^{- 1/2}k^{- (d+ \beta -1)}$, for $k> t^{-b}$. 
  
  We may immediately apply these standard (weak) turbulence results to the case of strong turbulence. Since for $\gamma>d$ the scaling $f(\xi) = \xi^{-\g}$ is outside of the convergence window, we take the effective interaction (\ref{BigLam}) with $\beta=2$. This gives $a=d/2$,  $b=1/2$, and
\be 
n_k\propto t^{-1/2}k^{-d-1} \label{ss1}
\ee 
for $k>t^{-1/2}$.  The scaling exponent  $\gamma=d{+}1$ in \eqref{ss1} is larger than the  weak turbulence scaling exponent, since the nonlinear effects suppress interactions. From the  kinetic equation with $\beta=2$ one can estimate the scaling of the interaction time, $n_k^{-1}dn_k/dt\propto k^{2(d-\gamma+1)}$, which is independent of $k$ for (\ref{ss1}), as it must be in a self-similar regime.  The scaling \eqref{ss1} was indeed observed in 3D numerics \cite{B15}: $n_k(t)=t^{-1/2}k^{-4}$,   in a widening interval between $t^{-1/2}<k<t^{-1/4}$. For $k> t^{-1/4}$, the quasi-steady state energy equipartition occurs. For $k<t^{-1/2}$  the plateau grows as $n_k\propto t^{d/2}$. Note that $\mL_-\simeq \lambda {\cal  N}k^{-2}$ is time-independent in this case.\\[-5pt]

\textit{Forced turbulence: }Let us now consider the formation of an inverse cascade of wave action under  pumping at high $k$, which provides for a linearly growing total number of waves. Naively, one would say that the front $k_0\propto t^{-1/2}$ leaves behind (at $k>k_0$) the steady constant-flux spectrum (\ref{ss2}). However, we have shown that (\ref{ss2}) is not a stationary solution. This implies the impossibility of a pumping-generated self-similar evolution for $n_k(t)$. Indeed, the effective interaction (\ref{BigLam}) depends on the total wave action $\mN$, which is now time-dependent. This means that as time progresses the spectrum changes at all $k$, and not just at the location of the moving front. An analytic description of such a complicated evolution is a challenge for future work. Numerics show a continuing yet slowing-down spectrum steepening upon propagation to lower $k$, accompanied by a flattening into an equipartition $n_k\propto k^{-2}$ at higher $k$, both in 2D and in 3D \cite{Pre,Naz23}. One could speculate that the slope of the spectrum asymptotically approaches $\gamma=d+2$ -- something future numerics would need to verify --  which could be natural if this is the edge of the convergence window of the momenta integrals  $p_2, p_3, p_4$ in $\tilde I(p_1)$ (\ref{largeNKE}). After $k_0(t)$ reaches the box size $L$,  a condensate starts growing with a steady equipartition for $kL\gg1$ \cite{OT,PT}.

\subsection{Focusing ($\lam<0$)}

We now consider the stationary inverse cascade in the focusing case.  It is natural to expect that as we move along the cascade to lower $k$, the change in the scaling exponent is monotonic: from the KZ scaling, $\g = d{-}2/3$ at large $k$, to a smaller exponent at lower $k$.
There is no IR divergence for less steep spectra. There also no UV divergence in the integral in $\mL_-$  if $\g>d{-}2$, as noted  after (\ref{ReLsmallq0}). The absence of either a UV or IR cutoff would lead to the scaling (\ref{ss2}), which is steeper than KZ (and, in any case, not a consistent stationary solution). Thus  the strong turbulence spectrum must have a scaling $\g \leq d{-}2$. 

In this case, we see from (\ref{Lminus03d}) and (\ref{ReLsmallq0}) that the integral for $\text{Re } \mL_-$   has a UV divergence. This comes with a built-in UV cutoff,  $k_*$, which is the scale at which the nonlinearity is of order-one and the spectrum starts transitioning to the KZ solution. The dominant contribution to $\mL_-$ (assuming $p_- \ll k_*$) therefore comes from region $p_-\ll q\ll k_*$.  Working in three dimensions, and assuming $\g <1$, 
\be \label{315}
\text{Re }\mL_- \approx  - 4\pi \lam \int_{p_-\ll q\ll k_*} dq\,  n_q \sim -\lam k_{*}^{1-\g}~,~ \ \ \ \ \g<1~.
\ee
If this were true, then the effective interaction (\ref{Lambda1234}) would behave as $
\Lambda_{1234} \sim k_*^{\g-1} $,  decreasing with an increase of the turbulence level  and $k_*=(Q\lambda)^{1/4}$. Such a decrease is inconsistent with our expectation of monotonicity: in (\ref{sgnA}) we found that the one-loop correction in the focusing case leads to an increase of the (angle-integrated) effective interaction. It would be odd if, as we move along the cascade, the effective interaction first increases and then decreases.

The only remaining option for the IR scaling exponent appears to be $\g = 1$. However, the pure power law $n_q = \frac{1}{4\pi |\lam| q}$ in (\ref{315}) leads to a loop integral that  grows with $k_*$:
$
\text{Re }\mL_- \approx  \log\frac{k_*}{p_-} + \ldots
$.
The ellipses represents  terms that scale as powers of $p_-/q$ and $q/k_*$, arising from corrections to the integrand at $q\simeq p_-$ and $q\simeq k_*$. The growth of $\text{Re }\mL_- $ again results in an unphysical decrease of the interaction $\Lambda_{1234}$, similar to what occurs with $\g<1$. A substantial increase of the interaction, either by moving deeper into the IR or by increasing $\lam$, can be achieved only by approaching the pole $\mL_- =1$ in the effective interaction:
\be \label{Lam316}
|\Lambda_{1234}|^2= \frac{\lam^2}{|1 - \mL_-|^2} =\frac{\lam^2}{(1{-} \text{Re } \mL_-)^2 + (\text{Im } \mL_-)^2}~.
\ee
If $\text{Re } \mL_- \approx 1$ and $\text{Im } \mL_-$ is small, then $\Lambda_{1234}$ can exceed the bare interaction value $\lam$.
This seems possible in the focusing case, since the angle-integrated vertex renormalization (\ref{sgnA}) was found to be positive in this case. As the nonlinearity becomes stronger, the angle averaged $\mL_-$ keeps increasing. If $\text{Im } \mL_-$  is small, then  $\text{Re } \mL_- \approx 1$ is the boundary of how large the effective nonlinearity parameter can be.  

With logarithmic accuracy, the $n_k$ that gives  $\text{Re } \mL_- \approx 1$ is, 
\be \label{nklog0}
n_k = \frac{1}{4\pi |\lam| \log \frac{k_*}{k}} \frac{1}{k}~,~ \ \ \ \ k\ll k_*~.
\ee
With this choice we have
\be \label{ReL318}
\text{Re }\mL_- \approx  - 4\pi \lam \int_{p_-}^{k_*} dq\, n_q  \approx 1~,
\ee
because we may approximate $\log\frac{k_*}{q}$ to be a constant, since $k_* \gg q$ for all $q$ in the integration range. 
If we do the integral more precisely, we will produce a double log term, 
\be
\int dq \frac{1}{q \log\frac{k_*}{q}} =  - \log \log \frac{k_*}{q}~,
\ee
which grows even slower than a log. If we wish, we can further correct $n_k$ with a double log term. We emphasize that the expression (\ref{nklog0}) is only valid to leading order in $k/k_*$. In other words, inside the log one can just as well replace the cutoff $k_*$ with, for instance, $k_*/2$. 

Let us now look at what this $n_k$  implies for the imaginary part of $\mL_-$. From (\ref{Lminus03d}) we have, 
 \be \label{ImLm}
 \text{Im }\mL_- =   - \frac{2 \pi^2 \lam}{p_-} \int_ \frac{|p_-^2-\o_-|}{2 p_- }^{ \frac{|p_-^2+\o_-|}{2 p_- }} dq\, q\, n_q  \approx\frac{\pi}{2 p_-} \int_ \frac{|p_-^2-\o_-|}{2 p_- }^{ \frac{|p_-^2+\o_-|}{2 p_- }} dq \frac{1}{\log \frac{k_*}{q}} \sim \frac{1}{ \log \frac{k_*}{p_-}}~.
 \ee
 In the second equality we made use of $n_k$  in (\ref{nklog0}). We are assuming that $\o_-$ is of the order of $p_-^2$ and that the integration range spans a region of the order of $p_-$.  Note that the $n_k$ in  (\ref{nklog0}) is only valid for $k\ll k_*$, so (\ref{ImLm}) is valid for $p_- \ll k_*$. As a result, in the final equality we can assume that the logarithm is essentially constant over the range of integration, replacing $q$ by something of the order of $p_-$. 
Inserting the real and imaginary parts of $\mL_-$, (\ref{ReL318}) and (\ref{ImLm}), into $\Lambda_{1234}$ (\ref{Lam316}) gives, 
\be \label{Lamlog}
|\Lambda_{1234}|^2\sim \(\lam \log\frac{k_*}{k}\)^2~,~ \ \ \ \ \ \ k \ll k_*~,
\ee
which becomes large in the IR (small $k/k_*$), as desired. If we now substitute (\ref{nklog0}) and (\ref{Lamlog}) into the standard kinetic equation, we obtain a flux that,  instead of being  constant (as is necessary for a stationary solution), decreases along the cascade:
\be
Q(q)\simeq |\Lambda_{1234}|^2\frac{n_q^3q^{3d}}{\omega_q}\sim q^4 \(\log\frac{k_*}{q}\)^2~.\label{FF}
\ee
Thus,  (\ref{nklog0}) is not a cascade solution of  the large-$N$ kinetic equation. So, what  is it?
One may observe that, up to a logarithmic factor, it is the  critical balance scaling \cite{Phil,Goldreich,Newell,NS}: the $n_k$ for which the nonlinearity parameter is of order $1$, $\eps_k\sim\lambda n_kk^{d}/\omega_k\simeq 1$, which gives, for general dimension $d$, 
\be \label{Uni}
n_k \sim \frac{1}{\lam}k^{2-d}~.
\ee
Physically, critical balance means that  the nonlinear and linear timescales are of the same order for all wavenumbers below $k_*$. Wave collapses (light self-focusing), which  occur when the interaction becomes comparable to the dispersion, limit the growth of the nonlinearity as  $k$ decreases and give a universal,  flux-independent spectrum for $q\ll k_*=(\lambda Q)^{1/4}$. It was suggested in \cite{OT} that each collapse (self-focusing) event  transfers wave action directly to large $k$, where  dissipation halts the self-focusing and burns the action. This mechanism results in flux suppression at low $k$. A similar flux loop was  observed numerically in stably stratified \cite{strat} and compressible \cite{loop} 2D turbulence, where shocks play a role analogous to collapses, transfering energy directly to large wavenumbers.

\subsection{Two dimensions} \label{sec2d}
The peculiarity of 2D weak turbulence lies in the fact that the KZ state \eqref{KZ} has the wrong sign for the flux (positive instead of negative, as required for an inverse cascade). Consequently,   the spectrum is instead  close to that of  thermal equilibrium, $n_q\simeq (Q/\lambda^2)^{1/3}(k^{-2}+k_0^2)$ \cite{ZLF, OT,Nonlocal}, across all scales. The analytic expression for $n_k$ in the weak nonlinearity regime is not known;  however, numerical results indicate \cite{Nonlocal} that it is less steep than the thermal $1/k^2$ scaling. Without the precise form of $n_k$ at weak nonlinearity, we cannot compute the higher order corrections to $n_k$ in the UV. 

It seems reasonable to assume that, like in the three dimensional case, the effective interaction increases/decreases for wave attraction/repulsion, which makes the occupation numbers lower/higher, respectively. Indeed, the loop integral diverges for $n_k\propto k^{-2}$ in the IR: taking $q\to 0$ and integrating over angles we obtain the same $-\lambda$ sign as in 3D. 
As a result, as we flow into the deep IR  (the strong turbulence regime) we expect $n_k$ to approach (\ref{ss3}) in the defocusing case, at least in the large $N$ limit. In the focusing case, $n_k$ should flow in the IR to something less steep than $k^{-2}$. The critical balance solution,  $n_k = \text{const.}$, where the constant is such that $\text{Re }\mL_- = 1$, satisfies this criterion. Note that $\text{Im }\mL_- = 0$ in this case. In fact, $n_k = \text{const.}$ is manifestly a stationary solution of the large $N$ kinetic equation (\ref{largeNKE}). Of course, $n_k=\text{const.}$ is a stationary solution of the standard kinetic equation valid in the UV; the nontrivial statement here is that it is a stationary solution in the IR, where nonlinearity is strong. 

The 2D data presented in Figure 7 in Appendix B of  \cite{Pre} is consistent with this picture. The spectra for the focusing and defocusing cases flow together in the weak-turbulence regime and then deviate in the IR. The former goes down to a constant, while the latter goes up, possibly approaching \eqref{ss2}, see the last panel of Figure 6 in  \cite{Pre}. 

One reason why critical balance may be more likely in 2D  than in 3D is that collapses are weak in 2D, which means that the kinetic and potential energy are comparable for such events, which corresponds to  critical balance. While   $n_k = A$ is a solution for any $A$ and  any $\lambda$,  our specific predictions are: i) it is realized in the inverse cascade of strong turbulence for negative $\lambda$, ii) the value $A\simeq 1/\lambda$ is independent of the action pumping rate.

\subsection{Position space correlators}
It is instructive to look at how our spectra,  (\ref{ss1}), (\ref{ss2}), (\ref{ss3}), and (\ref{Uni}), translate into laws of decorrelation in position space. Since all of these exponents contain $-d$, the $r$-space laws are independent of the  dimensionality of space. Free-decay involves a decrease of perturbations (vortices, fronts), while maintaining a constant level of action, $\overline{|\Psi|^2}={\cal  N}$. The spectrum \eqref{ss1} gives  $\langle|\psi(r)-\psi(0)|^2\rangle\simeq {\cal  N} k_0(t) r$. 
Such a linear dependence on separation could arise from the possibility of encountering a front, as seen in acoustic systems with shocks.
 The spectrum \eqref{ss2} gives $\langle|\psi(r)-\psi(0)|^2\rangle\simeq {\cal  N}(k_0 r)^{2}$, characteristic of a spatially smooth (finite-gradient) $\Psi(r)$. This law of decorrelation can be interpreted as being proportional to the number of  vortices (points in 2D and lines in 3D) within the distance $r$, where $k_0$ represents the mean distance between  pairs \cite{Nowak:2011sk,Pre}. The steady cascade \eqref{ss3} gives  $\langle|\psi(r)-\psi(0)|^2\rangle\simeq {\cal  N} (k_0 r)^{2/3}$. 
A much faster power-law decrease of correlations occurs for  \eqref{KZ} and \eqref{Uni}. For  \eqref{KZ}  we have  $\langle\psi(r)\psi^*(0)\rangle\simeq Q^{1/3}(\lambda r)^{-2/3}=(Q/\lambda)^{1/2}(k_*r)^{-2/3}$, while for \eqref{Uni}  $\langle\psi(r)\psi^*(0)\rangle\simeq 1/\lambda r^{2}=(Q/\lambda)^{1/2}(k_*r)^{-2}$. The latter spectrum corresponds to the focusing case if $1/k_*$ is allowed to be interpreted as the mean size of collapsing wave packets. 
\\[-8pt]
\section{Discussion} \label{sec:Dis}

We have shown that, as a weakly turbulent inverse  cascade proceeds from smaller to larger scales,  the effective interaction increases/decreases for the focusing ($\lam{<}0$)/defocusing ($\lam{>}0$) nonlinear Schr\"odinger equation. This results in the occupation numbers being, respectively, lower/higher than the weak-turbulence Kolmogorov-Zakharov scaling. To the best of our knowledge, no such statement has been made before, even though empirical observations of spectral steepening with  increasing nonlinearity can be found in \cite{Naz23} and  Figure 4 of \cite{Nonlocal}, which describes simulations of the defocusing case. For the nonlinear Schr\"odinger equation with a large $N$ number of fields, we argued that in the defocusing case, the spectrum in the UV (strong turbulence) is  (\ref{ss3}). Numerical simulations of the evolution in the defocusing case in three dimensions give an exponent somewhere  between $4$ and $5$ in 3D \cite{Naz23} and between $3$ and $4$ in 2D \cite{Pre}.  

It is instructive to compare the present findings with those of a large-$N$ model having an interaction that is strongly local in momentum space \cite{RSch}. The key difference is that in the strongly local model, the locality of interactions prevents any possible divergences. In the defocusing case, it exhibits the scaling (\ref{ss2}) in the strong turbulence regime, which is not realized here since it would lead to an IR divergence. In the focusing case, the strongly local model \cite{RSch} exhibits critical balance scaling in the strong turbulence regime; here, however, the status of this regime is unclear, at least in 3D.

The change in the strength of the effective interaction as one flows from the UV to the IR is reminiscent of  renormalization group flow in quantum field theory. There, a positive/negative beta function quantifies  the extent to which the interaction gets weaker/stronger as one moves into the IR, with quantum electrodynamics having a positive beta function and quantum chromodynamics having a negative beta function. The former is referred to as screening, and the latter as antiscreening (potentially leading to eventual confinement). Renormalization of interactions in the quantum field theory context is due to interactions with particle-antiparticle virtual pairs in the quantum vacuum. In our case, it is due to interactions with classical statistical fluctuations of the turbulent (far-from-equilibrium) state.

In Kolmogorov turbulence, we associate dependence on the largest scale $L=1/k_0$ with intermittency: the spectral density decreases as $L$ increases, since the same cascade is sustained by stronger and rarer fluctuations. In wave turbulence, lower spectral density \eqref{Uni} and intermittency occur in the focusing case, while dependence on $L$ occurs in the defocusing case, where \eqref{ss3} grows with $L$.

To conclude, we have made progress in the analytic description of strong-turbulence spectra  in the large-$N$ limit. Further progress   may require using quantities other than the occupation numbers  $n_k=\langle|\Psi_k|^2\rangle$. In the defocusing case, useful quantities  may include  correlations of over-condensate fluctuations. In the focusing case, useful quantities may include collapsing caverns as highly correlated multi-mode states,  analogous to hadrons in quantum chromodynamics.

\sss*{Acknowledgments} 
We thank  S.~Nazarenko, D.~Schubring, M.~Smolkin, N.~Vladimirova, and Ying Zhu for helpful discussions. VR is supported by NSF grant 2209116, by BSF grant 2022113, and by the ITS through a Simons grant. GF is supported by the Excellence Center at WIS, the Simons grant 617006, the ISF grant  146845, the NSF-BSF grant 2020765, and the EU Horizon grant No 873028.

\appendix 

\section{Next-to-leading order kinetic equation} \label{apA}

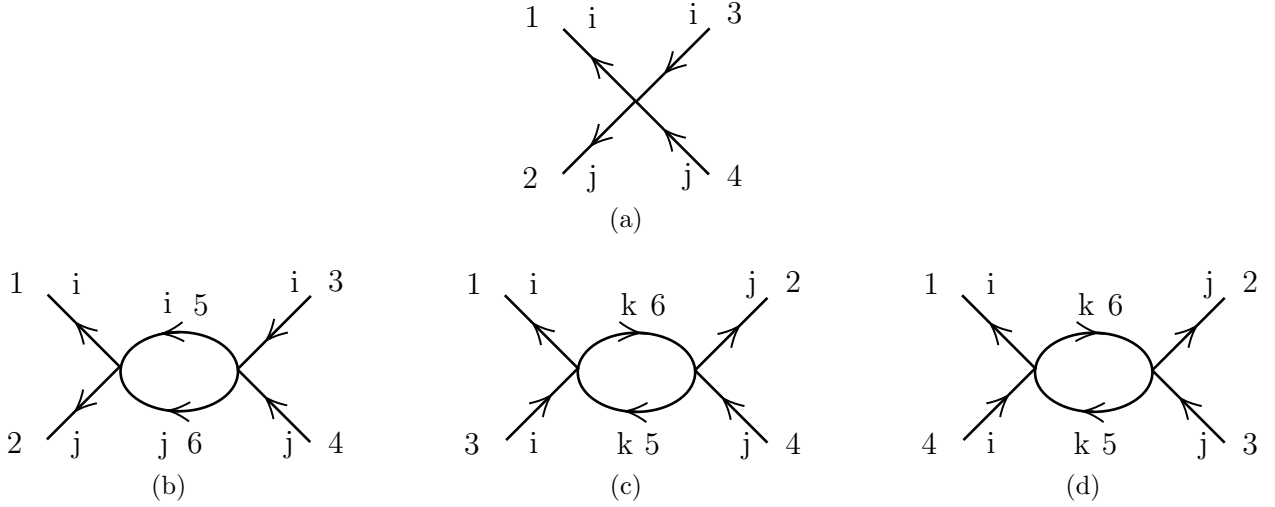
\begin{figure}[]
\centering
\subfloat[]{
\tikzset{every picture/.style={line width=1pt}} %set default line width to 0.75pt        
\begin{tikzpicture}[x=0.75pt,y=0.75pt,yscale=-1,xscale=1]
%Straight Lines [id:da9153477132713148] 
\draw    (120.99,79.59) -- (99.61,58.21) ;
\draw [shift={(98.2,56.8)}, rotate = 45] [color={rgb, 255:red, 0; green, 0; blue, 0 }  ][line width=0.75]    (10.93,-4.9) .. controls (6.95,-2.3) and (3.31,-0.67) .. (0,0) .. controls (3.31,0.67) and (6.95,2.3) .. (10.93,4.9)   ;
%Straight Lines [id:da3850884700046392] 
\draw    (84.2,42.8) -- (98.2,56.8) ;

%Straight Lines [id:da8337941395021127] 
\draw    (120.65,79.13) -- (99.28,100.51) ;
\draw [shift={(97.87,101.92)}, rotate = 315] [color={rgb, 255:red, 0; green, 0; blue, 0 }  ][line width=0.75]    (10.93,-4.9) .. controls (6.95,-2.3) and (3.31,-0.67) .. (0,0) .. controls (3.31,0.67) and (6.95,2.3) .. (10.93,4.9)   ;
%Straight Lines [id:da11003978097052458] 
\draw    (83.87,115.92) -- (97.87,101.92) ;

%Straight Lines [id:da7117654134297163] 
\draw    (157.65,116.25) -- (136.28,94.88) ;
\draw [shift={(134.87,93.47)}, rotate = 45] [color={rgb, 255:red, 0; green, 0; blue, 0 }  ][line width=0.75]    (10.93,-4.9) .. controls (6.95,-2.3) and (3.31,-0.67) .. (0,0) .. controls (3.31,0.67) and (6.95,2.3) .. (10.93,4.9)   ;
%Straight Lines [id:da7664461985067011] 
\draw    (120.87,79.47) -- (134.87,93.47) ;

%Straight Lines [id:da4927483440171442] 
\draw    (157.99,42.47) -- (136.61,63.84) ;
\draw [shift={(135.2,65.25)}, rotate = 315] [color={rgb, 255:red, 0; green, 0; blue, 0 }  ][line width=0.75]    (10.93,-4.9) .. controls (6.95,-2.3) and (3.31,-0.67) .. (0,0) .. controls (3.31,0.67) and (6.95,2.3) .. (10.93,4.9)   ;
%Straight Lines [id:da7078994304202018] 
\draw    (121.2,79.25) -- (135.2,65.25) ;
% Text Node
\draw (63,30) node [anchor=north west][inner sep=0.75pt]   [align=left] {1};
% Text Node
\draw (62,111) node [anchor=north west][inner sep=0.75pt]   [align=left] {2};
% Text Node
\draw (165,28) node [anchor=north west][inner sep=0.75pt]   [align=left] {3};
% Text Node
\draw (165,110) node [anchor=north west][inner sep=0.75pt]   [align=left] {4};
% Text Node
\draw (95,30) node [anchor=north west][inner sep=0.75pt]   [align=left] {i};
% Text Node
\draw (95,110) node [anchor=north west][inner sep=0.75pt]   [align=left] {j};
% Text Node
\draw (146,28) node [anchor=north west][inner sep=0.75pt]   [align=left] {i};
% Text Node
\draw (143,109) node [anchor=north west][inner sep=0.75pt]   [align=left] {j};
\end{tikzpicture}
}  \\
\subfloat[]{
\tikzset{every picture/.style={line width=1pt}} %set default line width to 0.75pt       
\begin{tikzpicture}[x=0.75pt,y=0.75pt,yscale=-1,xscale=1]
 

%Straight Lines [id:da8952577887886422] 
\draw    (120.99,79.59) -- (99.61,58.21) ;
\draw [shift={(98.2,56.8)}, rotate = 45] [color={rgb, 255:red, 0; green, 0; blue, 0 }  ][line width=0.75]    (10.93,-4.9) .. controls (6.95,-2.3) and (3.31,-0.67) .. (0,0) .. controls (3.31,0.67) and (6.95,2.3) .. (10.93,4.9)   ;
%Straight Lines [id:da41451408396593414] 
\draw    (84.2,42.8) -- (98.2,56.8) ;

%Straight Lines [id:da06431575564936653] 
\draw    (120.65,79.13) -- (99.28,100.51) ;
\draw [shift={(97.87,101.92)}, rotate = 315] [color={rgb, 255:red, 0; green, 0; blue, 0 }  ][line width=0.75]    (10.93,-4.9) .. controls (6.95,-2.3) and (3.31,-0.67) .. (0,0) .. controls (3.31,0.67) and (6.95,2.3) .. (10.93,4.9)   ;
%Straight Lines [id:da16889697375559964] 
\draw    (83.87,115.92) -- (97.87,101.92) ;

%Straight Lines [id:da1644669094129687] 
\draw    (216.65,117.25) -- (195.28,95.88) ;
\draw [shift={(193.87,94.47)}, rotate = 45] [color={rgb, 255:red, 0; green, 0; blue, 0 }  ][line width=0.75]    (10.93,-4.9) .. controls (6.95,-2.3) and (3.31,-0.67) .. (0,0) .. controls (3.31,0.67) and (6.95,2.3) .. (10.93,4.9)   ;
%Straight Lines [id:da2984933625391206] 
\draw    (179.87,80.47) -- (193.87,94.47) ;

%Straight Lines [id:da7589691045098796] 
\draw    (216.99,43.47) -- (195.61,64.84) ;
\draw [shift={(194.2,66.25)}, rotate = 315] [color={rgb, 255:red, 0; green, 0; blue, 0 }  ][line width=0.75]    (10.93,-4.9) .. controls (6.95,-2.3) and (3.31,-0.67) .. (0,0) .. controls (3.31,0.67) and (6.95,2.3) .. (10.93,4.9)   ;
%Straight Lines [id:da3022498007196087] 
\draw    (180.2,80.25) -- (194.2,66.25) ;

%Shape: Ellipse [id:dp3579924082553517] 
\draw   (120.99,81.71) .. controls (120.99,70.69) and (134.24,61.75) .. (150.59,61.75) .. controls (166.94,61.75) and (180.2,70.69) .. (180.2,81.71) .. controls (180.2,92.73) and (166.94,101.67) .. (150.59,101.67) .. controls (134.24,101.67) and (120.99,92.73) .. (120.99,81.71) -- cycle ;
%Straight Lines [id:da24563613243517846] 
\draw    (150.59,61.75) -- (143.78,62.37) ;
\draw [shift={(141.79,62.55)}, rotate = 354.81] [color={rgb, 255:red, 0; green, 0; blue, 0 }  ][line width=0.75]    (10.93,-4.9) .. controls (6.95,-2.3) and (3.31,-0.67) .. (0,0) .. controls (3.31,0.67) and (6.95,2.3) .. (10.93,4.9)   ;
%Straight Lines [id:da4328176872479478] 
\draw    (150.59,101.67) -- (146.93,101.45) ;
\draw [shift={(144.94,101.33)}, rotate = 3.37] [color={rgb, 255:red, 0; green, 0; blue, 0 }  ][line width=0.75]    (10.93,-4.9) .. controls (6.95,-2.3) and (3.31,-0.67) .. (0,0) .. controls (3.31,0.67) and (6.95,2.3) .. (10.93,4.9)   ;

% Text Node
\draw (63,30) node [anchor=north west][inner sep=0.75pt]   [align=left] {1};
% Text Node
\draw (62,111) node [anchor=north west][inner sep=0.75pt]   [align=left] {2};
% Text Node
\draw (224,29) node [anchor=north west][inner sep=0.75pt]   [align=left] {3};
% Text Node
\draw (224,111) node [anchor=north west][inner sep=0.75pt]   [align=left] {4};
% Text Node
\draw (95,30) node [anchor=north west][inner sep=0.75pt]   [align=left] {i};
% Text Node
\draw (95,110) node [anchor=north west][inner sep=0.75pt]   [align=left] {j};
% Text Node
\draw (205,29) node [anchor=north west][inner sep=0.75pt]   [align=left] {i};
% Text Node
\draw (202,110) node [anchor=north west][inner sep=0.75pt]   [align=left] {j};
% Text Node
\draw (141,40) node [anchor=north west][inner sep=0.75pt]   [align=left] {i};
% Text Node
\draw (138.94,110.33) node [anchor=north west][inner sep=0.75pt]   [align=left] {j};
% Text Node
\draw (156,41) node [anchor=north west][inner sep=0.75pt]   [align=left] {5};
% Text Node
\draw (153,111) node [anchor=north west][inner sep=0.75pt]   [align=left] {6};

\end{tikzpicture}
} \ \ \ \ \ \ \ \
\subfloat[]{

\tikzset{every picture/.style={line width=1pt}} %set default line width to 0.75pt        

\begin{tikzpicture}[x=0.75pt,y=0.75pt,yscale=-1,xscale=1]

%Straight Lines [id:da6536284336590911] 
\draw    (120.99,79.59) -- (99.61,58.21) ;
\draw [shift={(98.2,56.8)}, rotate = 45] [color={rgb, 255:red, 0; green, 0; blue, 0 }  ][line width=0.75]    (10.93,-4.9) .. controls (6.95,-2.3) and (3.31,-0.67) .. (0,0) .. controls (3.31,0.67) and (6.95,2.3) .. (10.93,4.9)   ;
%Straight Lines [id:da9982534738877924] 
\draw    (84.2,42.8) -- (98.2,56.8) ;

%Straight Lines [id:da1798845392114402] 
\draw    (84.87,115.92) -- (106.24,94.55) ;
\draw [shift={(107.65,93.13)}, rotate = 135] [color={rgb, 255:red, 0; green, 0; blue, 0 }  ][line width=0.75]    (10.93,-4.9) .. controls (6.95,-2.3) and (3.31,-0.67) .. (0,0) .. controls (3.31,0.67) and (6.95,2.3) .. (10.93,4.9)   ;
%Straight Lines [id:da4073203382472357] 
\draw    (121.65,79.13) -- (107.65,93.13) ;

%Straight Lines [id:da3010890665331972] 
\draw    (216.65,117.25) -- (195.28,95.88) ;
\draw [shift={(193.87,94.47)}, rotate = 45] [color={rgb, 255:red, 0; green, 0; blue, 0 }  ][line width=0.75]    (10.93,-4.9) .. controls (6.95,-2.3) and (3.31,-0.67) .. (0,0) .. controls (3.31,0.67) and (6.95,2.3) .. (10.93,4.9)   ;
%Straight Lines [id:da07001653100141458] 
\draw    (179.87,80.47) -- (193.87,94.47) ;

%Shape: Ellipse [id:dp37207364622839356] 
\draw   (120.99,81.71) .. controls (120.99,70.69) and (134.24,61.75) .. (150.59,61.75) .. controls (166.94,61.75) and (180.2,70.69) .. (180.2,81.71) .. controls (180.2,92.73) and (166.94,101.67) .. (150.59,101.67) .. controls (134.24,101.67) and (120.99,92.73) .. (120.99,81.71) -- cycle ;
%Straight Lines [id:da7816491313522526] 
\draw    (156.11,62.11) -- (152.59,61.88) ;
\draw [shift={(152.59,61.88)}, rotate = 183.71] [color={rgb, 255:red, 0; green, 0; blue, 0 }  ][line width=0.75]    (10.93,-4.9) .. controls (6.95,-2.3) and (3.31,-0.67) .. (0,0) .. controls (3.31,0.67) and (6.95,2.3) .. (10.93,4.9)   ;
%Straight Lines [id:da6992984700470459] 
\draw    (150.59,101.67) -- (146.93,101.45) ;
\draw [shift={(144.94,101.33)}, rotate = 3.37] [color={rgb, 255:red, 0; green, 0; blue, 0 }  ][line width=0.75]    (10.93,-4.9) .. controls (6.95,-2.3) and (3.31,-0.67) .. (0,0) .. controls (3.31,0.67) and (6.95,2.3) .. (10.93,4.9)   ;
%Straight Lines [id:da8812714453372484] 
\draw    (179.87,80.92) -- (201.24,59.55) ;
\draw [shift={(202.65,58.13)}, rotate = 135] [color={rgb, 255:red, 0; green, 0; blue, 0 }  ][line width=0.75]    (10.93,-4.9) .. controls (6.95,-2.3) and (3.31,-0.67) .. (0,0) .. controls (3.31,0.67) and (6.95,2.3) .. (10.93,4.9)   ;
%Straight Lines [id:da5343707307385305] 
\draw    (216.65,44.13) -- (202.65,58.13) ;

% Text Node
\draw (63,30) node [anchor=north west][inner sep=0.75pt]   [align=left] {1};
% Text Node
\draw (62,111) node [anchor=north west][inner sep=0.75pt]   [align=left] {3};
% Text Node
\draw (224,29) node [anchor=north west][inner sep=0.75pt]   [align=left] {2};
% Text Node
\draw (224,111) node [anchor=north west][inner sep=0.75pt]   [align=left] {4};
% Text Node
\draw (95,30) node [anchor=north west][inner sep=0.75pt]   [align=left] {i};
% Text Node
\draw (95,110) node [anchor=north west][inner sep=0.75pt]   [align=left] {i};
% Text Node
\draw (205,29) node [anchor=north west][inner sep=0.75pt]   [align=left] {j};
% Text Node
\draw (202,110) node [anchor=north west][inner sep=0.75pt]   [align=left] {j};
% Text Node
\draw (141,40) node [anchor=north west][inner sep=0.75pt]   [align=left] {k};
% Text Node
\draw (138.94,110.33) node [anchor=north west][inner sep=0.75pt]   [align=left] {k};
% Text Node
\draw (156,41) node [anchor=north west][inner sep=0.75pt]   [align=left] {6};
% Text Node
\draw (153,111) node [anchor=north west][inner sep=0.75pt]   [align=left] {5};
\end{tikzpicture}
} \ \ \ \ \ \ \ \
\subfloat[]{
\tikzset{every picture/.style={line width=1pt}} %set default line width to 0.75pt       
\begin{tikzpicture}[x=0.75pt,y=0.75pt,yscale=-1,xscale=1]

%Straight Lines [id:da8494672795448038] 
\draw    (120.99,79.59) -- (99.61,58.21) ;
\draw [shift={(98.2,56.8)}, rotate = 45] [color={rgb, 255:red, 0; green, 0; blue, 0 }  ][line width=0.75]    (10.93,-4.9) .. controls (6.95,-2.3) and (3.31,-0.67) .. (0,0) .. controls (3.31,0.67) and (6.95,2.3) .. (10.93,4.9)   ;
%Straight Lines [id:da5602988818023655] 
\draw    (84.2,42.8) -- (98.2,56.8) ;

%Straight Lines [id:da6235991143016453] 
\draw    (84.87,115.92) -- (106.24,94.55) ;
\draw [shift={(107.65,93.13)}, rotate = 135] [color={rgb, 255:red, 0; green, 0; blue, 0 }  ][line width=0.75]    (10.93,-4.9) .. controls (6.95,-2.3) and (3.31,-0.67) .. (0,0) .. controls (3.31,0.67) and (6.95,2.3) .. (10.93,4.9)   ;
%Straight Lines [id:da7109278035111258] 
\draw    (121.65,79.13) -- (107.65,93.13) ;

%Straight Lines [id:da08889333902350072] 
\draw    (216.65,117.25) -- (195.28,95.88) ;
\draw [shift={(193.87,94.47)}, rotate = 45] [color={rgb, 255:red, 0; green, 0; blue, 0 }  ][line width=0.75]    (10.93,-4.9) .. controls (6.95,-2.3) and (3.31,-0.67) .. (0,0) .. controls (3.31,0.67) and (6.95,2.3) .. (10.93,4.9)   ;
%Straight Lines [id:da7255617736761344] 
\draw    (179.87,80.47) -- (193.87,94.47) ;

%Shape: Ellipse [id:dp5121988410974818] 
\draw   (120.99,81.71) .. controls (120.99,70.69) and (134.24,61.75) .. (150.59,61.75) .. controls (166.94,61.75) and (180.2,70.69) .. (180.2,81.71) .. controls (180.2,92.73) and (166.94,101.67) .. (150.59,101.67) .. controls (134.24,101.67) and (120.99,92.73) .. (120.99,81.71) -- cycle ;
%Straight Lines [id:da40175426808736503] 
\draw    (156.11,62.11) -- (152.59,61.88) ;
\draw [shift={(152.59,61.88)}, rotate = 183.71] [color={rgb, 255:red, 0; green, 0; blue, 0 }  ][line width=0.75]    (10.93,-4.9) .. controls (6.95,-2.3) and (3.31,-0.67) .. (0,0) .. controls (3.31,0.67) and (6.95,2.3) .. (10.93,4.9)   ;
%Straight Lines [id:da25493545481358637] 
\draw    (150.59,101.67) -- (146.93,101.45) ;
\draw [shift={(144.94,101.33)}, rotate = 3.37] [color={rgb, 255:red, 0; green, 0; blue, 0 }  ][line width=0.75]    (10.93,-4.9) .. controls (6.95,-2.3) and (3.31,-0.67) .. (0,0) .. controls (3.31,0.67) and (6.95,2.3) .. (10.93,4.9)   ;
%Straight Lines [id:da4760331387646788] 
\draw    (179.87,80.92) -- (201.24,59.55) ;
\draw [shift={(202.65,58.13)}, rotate = 135] [color={rgb, 255:red, 0; green, 0; blue, 0 }  ][line width=0.75]    (10.93,-4.9) .. controls (6.95,-2.3) and (3.31,-0.67) .. (0,0) .. controls (3.31,0.67) and (6.95,2.3) .. (10.93,4.9)   ;
%Straight Lines [id:da5985173041822004] 
\draw    (216.65,44.13) -- (202.65,58.13) ;

% Text Node
\draw (63,30) node [anchor=north west][inner sep=0.75pt]   [align=left] {1};
% Text Node
\draw (62,111) node [anchor=north west][inner sep=0.75pt]   [align=left] {4};
% Text Node
\draw (224,29) node [anchor=north west][inner sep=0.75pt]   [align=left] {2};
% Text Node
\draw (224,111) node [anchor=north west][inner sep=0.75pt]   [align=left] {3};
% Text Node
\draw (95,30) node [anchor=north west][inner sep=0.75pt]   [align=left] {i};
% Text Node
\draw (95,110) node [anchor=north west][inner sep=0.75pt]   [align=left] {i};
% Text Node
\draw (205,29) node [anchor=north west][inner sep=0.75pt]   [align=left] {j};
% Text Node
\draw (202,110) node [anchor=north west][inner sep=0.75pt]   [align=left] {j};
% Text Node
\draw (141,40) node [anchor=north west][inner sep=0.75pt]   [align=left] {k};
% Text Node
\draw (138.94,110.33) node [anchor=north west][inner sep=0.75pt]   [align=left] {k};
% Text Node
\draw (156,41) node [anchor=north west][inner sep=0.75pt]   [align=left] {6};
% Text Node
\draw (153,111) node [anchor=north west][inner sep=0.75pt]   [align=left] {5};
\end{tikzpicture}
}
\caption{ Tree-level and  one-loop Feynman diagrams contributing to the kinetic equation.}\label{tree}
\end{figure}

In this appendix we derive several of the results quoted in Sec.~\ref{sec2}. 
The standard wave kinetic equation for the nonlinear Schr\"odinger equation is given by, 
\be \label{KE1}
\frac{\d n_1}{\d t} = 16\pi\lam^2 \int d\v p_2 d\v p_3 d\v p_4\,  n_1 n_2 n_3 n_4 \Big( \frac{1}{n_1} {+} \frac{1}{n_2}{-}\frac{1}{n_3} {-} \frac{1}{n_4} \Big)  \delta(\o_{p_1p_2; p_3 p_4})\delta(\v p_{12;34})~,
\ee
where $\o_p = p^2$ is the frequency,  $\o_{p_1p_2; p_3 p_4}\equiv \o_{p_1}{+}\o_{p_2}{-} \o_{p_3}{-}\o_{ p_4}$, $\v p_{12;34}\equiv \v p_1 {+}\v p_2{-}\v p_3 {-}\v p_4$, and $n_i \equiv n_{p_i}$.  This is valid at leading order in  $\lam$. The kinetic equation at next-to-leading order in $\lam$ is \cite{RS1, RS2}:
\be \label{12}
\frac{\d n_1}{\d t} = 16\pi \lam^2 \text{Re} \int d\v p_2 d\v p_3 d\v p_4\,  n_1 n_2 n_3 n_4\Big( \frac{1}{n_1} {+} \frac{1}{n_2}{-}\frac{1}{n_3} {-} \frac{1}{n_4} \Big)\\
\(1 + 2\mL_+ + 8\mL_-\) \delta(\o_{p_1p_2; p_3 p_4})\delta(\v p_{12;34})~,
\ee
where
\be \label{A2}
\mL_+ =  2\lam \int d \v p_5 d\v p_6\frac{n_{p_5} {+} n_{p_6}}{\o_{p_1 p_2;p_5 p_6}{+}i\eps}\delta(\v p_{12;56})~, \ \ \ \ \ \mL_- = 2\lam  \int d \v p_5 d\v p_6 \frac{n_{p_6} {-} n_{p_5}}{\o_{p_1 p_6;p_3 p_5}{+}i\eps}\delta(\v p_{16;35})~.
\ee
 Diagrammatically, the leading order term is represented by the tree diagram in Fig.~\ref{tree}(a), while  the one-loop diagram Fig.~\ref{tree}(b) gives the $\mL_+$ contribution, and the other one loop diagrams, Fig.~\ref{tree}(c) and Fig.~\ref{tree}(d), give the $\mL_-$ contribution. For the moment one should ignore the $i,j,k$ indices in the diagram. In principle, one can compute the kinetic equation to arbitrary order in $\lam$ \cite{RSSS}, but the higher the order, the more terms there are and the more unwieldy the expression. 

%%%%%%%%
%%%%% START FIGURE
%%%%%%%%
%%%%% End FIGURE

Let us write the expressions for $\mL_{\pm}$ more explicitly. We define
\bea \label{114}
\omega_+ &=& \o_{p_1} + \o_{p_2}~\ \ \ \ \ \ \o_- = \o_{p_4} - \o_{p_2} \\  \nn
 \v p_+ &=& \vec{p}_1 + \v p_2~,  \ \  \ \ \  \ \ \  \v p_- = \v p_4 - \v p_2~.
\eea
The following identities will be useful later on:
\bea \nn
\frac{p_+^2 - \o_+}{2} &=& p_1{\cdot} p_2~, \ \ \ \ \ \ \ \ \ \  \ p_-^2{+}\o_- = 2\v p_4 {\cdot} (\v p_4 {-}\v p_2)~, \ \ \ \ \\ \label{A5id}
2\o_+{ -} p_+^2 &=& (\v p_1 {-}\v p_2)^2~,  \ \ \ \ \ \  p_-^2{-}\o_- = 2\v p_2 {\cdot} (\v p_2 {-}\v p_4)~.
\eea
One can see that $\mL_+$ depends only on $\o_+$ and $\v p_+$, while $\mL_-$ depends only on $\o_-$ and $\v p_-$. We have, 
\bea \label{mLp1}
 \mL_+ &=& 4\lam\int d^d q \frac{n_{q}}{\o_+ -q^2 -(\v p_+{ -} \v q)^2 {+}i\eps}~, \\ \label{mLm1}
 \mL_- &=& -2\lam \int d^d q\, n_q\[\frac{1}{\o_- {-} q^2 + (\v p_- {-} \v q)^2{+}i\eps }-\frac{1}{\o_- {+}q^2 {-} (\v p_- {+ }\v q)^2 {+}i\eps}\]~,
 \eea
 where we  used that $\o_p = p^2$ and are denoting $p \equiv |\v p|$.
It will be convenient to write $\mL_-$ as
\be \label{mLm2}
\mL_- = \mI(\o_-) + \mI^*(-\o_-)~, \ \ \ \ \ \mI(\o_-) = -2 \lam \int d^d q\, n_q \frac{1}{\o_- {-} q^2 + (\v p_- {-} \v q)^2 +i\eps}~,
\ee
which we obtained from (\ref{mLm1}) by changing variables $\v q \rightarrow - \v q$ in the second term. 

\subsection{Renormalized Interaction}
  Let us go back to the one-loop kinetic equation (\ref{12}). We rewrite it as
  \bea \label{12v2}
\frac{\d n_1}{\d t} &=& 16\pi \, \int d\v p_2 d\v p_3 d\v p_4\,  n_1 n_2 n_3 n_4\,  \Lambda_{1234}^2\, \Big( \frac{1}{n_1} {+} \frac{1}{n_2}{-}\frac{1}{n_3} {-} \frac{1}{n_4} \Big)  \delta(\o_{p_1p_2; p_3 p_4})\delta(\v p_{12;34}) \\ \nn
\Lambda_{1234}^2 &=& \lam^2 \(1 + 2\mL_+ + 8\mL_-\)
\eea
so that it looks identical to a tree-level kinetic equation, but with an effective (or renormalized) squared interaction $\Lambda_{1234}^2$. 

We would like a slightly better form, so that it is manifest that $\Lambda_{1234}^2$, multiplied by the momentum and frequency conserving delta functions, enjoys all the symmetries of an interaction. Namely, in a Hamiltonian with a general quartic interaction term $\sum_{p_i} \Lambda_{1234} \Psi_1^* \Psi_2^* \Psi_3 \Psi_4$, the interaction $\Lambda_{1234}$ must have the symmetries $\Lambda_{1234} = \Lambda_{2134} = \Lambda_{1243} = \Lambda_{3412}^*$.    From the form of $\mL_+$ in (\ref{A2}) it is clear that $\mL_+$ is symmetric under $p_1 \leftrightarrow p_2$, and the symmetry under $p_1 \leftrightarrow p_3, p_2\leftrightarrow p_4$ is an immediate consequence of the frequency conserving delta function $\delta(\o_{p_1p_2; p_3 p_4})$ in the kinetic equation.~\footnote{These symmetries are less immediate from the explicit form of $\mL_+$ in (\ref{Lplus2}), but still clearly there, as one can see by writing $|\v p_1 - \v p_2| = \sqrt{2\o_+ - p_+^2}$.} Looking at $\mL_-$ in (\ref{A2}), under the exchange $p_1 \leftrightarrow p_3, p_2\leftrightarrow p_4$  it transforms into, 
\be
\mL_ - \rightarrow 2\lam \int d \v p_5 d\v p_6\   \frac{n_{p_6} {-} n_{p_5}}{\o_{p_3 p_6;p_1 p_5}{+}i\eps}\delta(\v p_{16;35})= 2\lam \int d \v p_5 d\v p_6\  \frac{n_{p_6} {-} n_{p_5}}{\o_{p_1 p_6;p_3 p_5}{-}i\eps}\delta(\v p_{16;35}) = \mL_-^*~,
\ee
where in the second equality we changed dummy variables $p_5 \leftrightarrow p_6$. The symmetry of  the $\mL_-$ piece  under $p_3 \leftrightarrow p_4$ is evident, because both $p_3$ and $p_4$ are integrated over. To show the symmetry of the $\mL_-$ piece under $p_1 \leftrightarrow p_2$ one uses that this transforms the denominator $\o_{16;35}$ in $\mL_-$ into $\o_{26;35}$, which after exchanging $p_3 \leftrightarrow p_4$ (since they are both integrated over in the kinetic equation) and exchanging $p_5 \leftrightarrow p_6$  and using the frequency conserving delta function, gives back $\mL_-^*$. 

Thus, referring to $\mL_{\pm}$ in (\ref{A2}) as $\mL_{\pm}(1,2,3,4)$ we may write $\Lambda_{1234}^2$ in a way that is manifestly symmetric, 
\bml \label{Lam1234}
\Lambda_{1234}^2  = \lam^2 + \lam^2(\mL_+(1,2,3,4)+\mL_+(3,4,1,2)) 
+ \frac{4}{3}\lam^2\Big(\mL_-(1,2,3,4)+ \mL_-(2,1,3,4) +\mL_-(1,2,4,3)\\+\mL_-(3,4,1,2)+ \mL_-(4,3,1,2)+ \mL_-(3,4,2,1)\Big)~.
\end{multline}

\subsection{Angular integrals}
Assuming that the spectrum $n_k$ is independent of the direction of $\v k$, we many evaluate the angular integrals appearing in $\mL_+$ and $\mL_-$. This is dimension-dependent; we separately consider two and three dimensions, starting with three dimensions.

\subsubsection{Three dimensions}
Writing $\v p {\cdot} \v q = p q \cos \theta$ and specializing to $d=3$,  the real part of $\mL_+$ (\ref{mLp1}) becomes,
\bea \label{Lplus} 
\!\!\!\text{Re } \mL_+ \!\!\!&=&\!\!\!\!\frac{4\pi \lam}{p_+}\! \int dq\, q\,  d (\cos \theta) \frac{n_q}{\cos \theta + \frac{\o_+ - 2 q^2 - p_+^2}{2 p_+q}} =\frac{4\pi \lam}{p_+} \int_0^{\infty}\!\!\! dq\, q\, n_q \log\Big|\frac{\o_+{ -} q^2{ -} (q{-}p_+)^2}{\o_+ {-} q^2 {-} (q{+}p_+)^2}\Big|\\ \nn
&=&\!\!\!\frac{4\pi \lam}{p_+}\! \int_0^{\infty}\!\! dq\, q\, n_q \log\Big|\frac{q^2 {-} q p_+ {+} \frac{p_+^2- \o_+}{2}}{q^2 {+} q p_+ {+} \frac{p_+^2- \o_+}{2}}\Big|
= \frac{4\pi \lam}{|\v p_1{+}\v p_2|} \int_0^{\infty} dq\, q\, n_q \log\Big|\frac{q^2 {-} q |\v p_1{+}\v p_2| {+}\v p_1{\cdot} \v p_2}{q^2 {+} q |\v p_1{+}\v p_2|{+}\v p_1{\cdot} \v p_2}\Big| ~,
\eea
where we performed the angular integral by making use of the  identity 
\be
\int_{-1}^1 dy \frac{1}{y+\al} = \log\Big|\frac{1+\al}{1-\al}\Big|~,
\ee
and wrote the result in several alternate forms that will be convenient later on. 
Looking now at $\mL_-$ (\ref{mLm2}) we have
\bea \label{A20}
\mI(\o_-) &=& \frac{2\pi \lam}{p_-}\int dq\, q\,n_q\,  d(\cos \theta) \frac{1}{\cos \theta - \frac{\o_- +p_-^2}{2p_- q}{-}i\eps}\\
&=& \frac{2\pi \lam}{p_-} \int dq\, q\, n_q \(\log\Big| \frac{\o_- {-}q^2{+}(p_- {-}q)^2}{\o_- {-}q^2{+}(p_- {+}q)^2}\Big| + i \pi \Theta\(1- \frac{|p_-^2+\o_-|}{2 p_- q}\) \)~,
\eea
where $\Theta(x)$ is the step theta function, and we evaluated the $\theta$ integral by  
 splitting the fraction into a principal part and a delta function,
 \be
 \frac{1}{x{-} a {-} i\eps} = P \frac{1}{x {-} a} + i\pi \delta(x{-}a)~.
 \ee
Thus, we get, 
 \be \label{Lminus0}
 \mL_- =   \frac{2\pi \lam}{p_-} \int_0^{\infty} dq\, q\, n_q \log\Big| \frac{((p_-{-}q)^2{-}q^2)^2 {-}\o_-^2}{((p_-{+}q)^2{-}q^2)^2 {-}\o_-^2}\Big|  - i\frac{2 \pi^2 \lam}{p_-} \int_ \frac{|p_-^2-\o_-|}{2 p_- }^{ \frac{|p_-^2+\o_-|}{2 p_- }} dq\, q\, n_q~
 \ee
 which is the form quoted in the main body, (\ref{Lminus03d}). 
Only the real part of $\mL_-$ is relevant for the next-to-leading order kinetic equation. However, for the large $N$ (all orders in $\lam$) kinetic equation (\ref{largeNKE}), both the real and imaginary parts of $\mL_-$ appear. We may equivalently rewrite the real part as, 
\bea
\text{Re } \mL_- &=&   \frac{2\pi \lam}{p_-} \int_0^{\infty} dq\, q\, n_q \log\Big| \frac{q^2 - q p_- + \frac{p_-^4 - \o_-^2}{4 p_-^2}}{q^2 +q p_- + \frac{p_-^4 - \o_-^2}{4 p_-^2}}\Big|  \\
&=& \frac{2\pi \lam}{|\v p_2{-}\v p_4|} \int_0^{\infty} dq\, q\, n_q \log\Big| \frac{q^2 - q |\v p_2{-}\v p_4| - \frac{\v p_2{\cdot}(\v p_2{-}\v p_4)\v p_4{\cdot}(\v p_2{-}\v p_4)}{(\v p_2 {-}\v p_4)^2}}{q^2 + q |\v p_2{-}\v p_4| - \frac{\v p_2{\cdot}(\v p_2{-}\v p_4)\v p_4{\cdot}(\v p_2{-}\v p_4)}{(\v p_2 {-}\v p_4)^2}}\Big|~,
\eea
which is the form quoted in the main body in (\ref{loop}). 

\subsubsection{Two dimensions}
Specializing to $d=2$ we have for the real part of $\mL_+$ (\ref{mLp1}),
 \bea \label{ReLp2d}
\text{Re } \mL_+ &=&\frac{2\lam}{p_+}\int dq d \theta \frac{ n_q}{\cos \theta + \frac{\o_+ {-} 2 q^2 {-} p_+^2}{2 p_+q}}  \\ \nn
&= & 8\pi \lam\int_0^{\infty}\!\! d q\, q\, n_q  \frac{ \text{sgn}(\o_+ {-} 2q^2 {-} p_+^2)}{\sqrt{(\o_+ {-} 2 q^2 {-} p_+^2)^2-(2p_+ q)^2}}  \Theta((\o_+ {-} 2 q^2 {-} p_+^2)^2>(2p_+ q)^2)~,
\eea
where $\sgn(x)$ is $1$ for positive $x$,  $-1$ for negative $x$, and zero for $x=0$, and
 we performed the angular integral through use of the identity, 
 \be
\int_0^{2\pi}d\theta \frac{1}{\cos \theta -a-i\epsilon} =  - \frac{2\pi\ \text{sgn}(a) }{\sqrt{a^2{-}1}}\Theta(|a|>1) +\frac{2 i \pi}{\sqrt{1{-}a^2}}\Theta(|a|<1)~,
 \ee
which can be found by contour integration (one can change variables to $z=e^{i \theta}$). Looking now at $\mL_-$ (\ref{mLm2}) we have
\small
 \bea \nn
 \!\!\!\!\!\!\!\!\!\!\!\!\!\!\mI(\o_-) =  \frac{ \lam}{p_-}\int dq\, d\theta \frac{n_q}{\cos \theta - \frac{\o_- +p_-^2}{2p_- q}{-}i\eps}   
&=&-4\pi \lam \int_{0}^{\frac{|p_-^2{+}\o_-|}{2 p_-}}\!\! dq\frac{  q\, n_q\,   \text{sgn}(p_-^2{+}\o_-)}{\sqrt{(p_-^2{+}\o_-)^2-(2p_- q)^2}} \\
 &+&   4\pi i \lam\int_{\frac{|p_-^2{+}\o_-|}{2 p_-}}^{\infty} dq \frac{ q\,n_q}{\sqrt{(2p_- q)^2-(p_-^2{+}\o_-)^2}}~.
\eea
\normalsize
Thus we get for $\mL_-$,
\small
 \bea \nn
\!\!\!\!\!\!\!\!\!\!\!\!\text{Re } \mL_- \!\!\!&=&\!\!\!  -4\pi \lam \int_{0}^{\frac{|p_-^2{+}\o_-|}{2 p_-}} \!\!\ dq\frac{q\,  n_q\,   \text{sgn}(p_-^2{+}\o_-)}{\sqrt{(p_-^2{+}\o_-)^2-(2p_- q)^2}}  -4\pi \lam \!\!\int_{0}^{\frac{|p_-^2{-}\o_-|}{2 p_-}}\!\! dq\frac{ q\, n_q\,  \text{sgn}(p_-^2{-}\o_-)}{\sqrt{(p_-^2{-}\o_-)^2-(2p_- q)^2}}  \\ \label{A30}
\!\!\!\!\!\!\!\!\!\!\!\!\text{Im } \mL_-\!\!\!&=&\!\!\!- 4\pi \lam\int_{\frac{|p_-^2{+}\o_-|}{2 p_-}}^{\infty} dq \frac{ q\, n_q}{\sqrt{(2p_- q)^2-(p_-^2{+}\o_-)^2}}+4\pi \lam\int_{\frac{|p_-^2{-}\o_-|}{2 p_-}}^{\infty} dq \frac{ q\, n_q}{\sqrt{(2p_- q)^2-(p_-^2{-}\o_-)^2}}~.\, \, \, \, \, \, \, 
 \eea
 \normalsize
Note that   $\text{Im } \mL_-=0$ for $n_q=$const, and Re $\mL_-=1$ for the particular constant, $n_q=-1/4\pi\lambda$, which makes sense only for $\lambda<0$.

\subsection{Next-to-leading order kinetic equation near the KZ state}
The next-to-leading order kinetic equation was given in (\ref{12}). If we are near the Kolmogorov-Zakharov state then at this order we may replace the $n_k$ within the loop integrals $\mL_{\pm}$ by the KZ solution $n_k = k^{-\g}$, and  carry out the integration over the loop momentum. We now do this explicitly in three dimensions.

There are several combinations of momenta and frequencies that will appear. In particular, we define
\be \label{tpm}
\t p_{\pm} = |\v p_1 + \v p_2 | \pm |\v p_1 - \v p_2|~.
\ee
Some identities that will be useful later on, in addition to (\ref{A5id}), are: 
\bea \nn
\t p_{\pm}^2 = 2 \o_+ \pm 2 p_+\sqrt{2\o_+ - p_+^2}~, \  \ \ \ \ \ \t p_+ \t p_- =2(p_+^2-\o_+)~, \ \ \ \  \ \t p_+^2 + \t p_-^2 = 4 \o_+\\
(\o_+ {-} 2 q^2 {-} p_+^2)^2-(2p_+ q)^2 = 4 q^4 - 4q^2 \o_+ + (p_+^2{-}\o_+)^2 = \frac{1}{4}(4 q^2 - {\t p_+}^2)(4 q^2 - {\t p_-}^2)~. \label{idenA27}
\eea

In three dimensions the inverse KZ cascade is $n_k = k^{-7/3}$. We insert this into $\text{Re }\mL_+$ (\ref{Lplus}) and evaluate the integral over the magnitude $q$ of the loop momentum to get, 
\be \label{Lplus2}
\text{Re }\mL_+ =\frac{4\pi \lam}{p_+} \int_0^{\infty} dq\, q^{-4/3} \log \Big| \frac{ (2q - \t p_+)(2q- \t p_-)}{(2q + \t p_+)(2q+ \t p_-)}\Big|= -\frac{4\pi^2 2^{\frac{1}{3}}\sqrt{3} \lam}{p_+} \( \frac{1}{\t p_+^{\, 1/3}}+ \frac{\sgn(\t p_-) }{|\t p_-|^{1/3}}\)~,
\ee
where  we made use of the integrals, 
\be
\int_0^{\infty} dx\,  x^{\mu-1} \log|1 {-} x| = \frac{\pi}{\mu} \cot (\pi \mu)~, \ \ \ \ \ \int_0^{\infty} dx\,  x^{\mu-1} \log(1 + x) = \frac{\pi}{\mu \sin \pi \mu}~,
\ee
valid for $-1<\text{Re}\, \mu<0$.  Likewise, we find the integral for $\text{Re }\mL_-$ (\ref{Lminus0}) evaluates to, 
 \bea \label{Lminus}
 \text{Re }\mL_- &=&- \frac{\pi^2 2^{4/3}\sqrt{3} \lam}{p_-^{2/3}}\( \frac{\sgn(\o_-{+}p_-^2)}{|\o_-{+}p_-^2|^{1/3} }- \frac{\sgn(\o_-{-}p_-^2)}{ |\o_-{-}p_-^2|^{1/3}}  \)\\ \label{Lminus2}
  &=& - \frac{\pi^2 2\sqrt{3} \lam}{|\v p_2{-}\v p_4|^{2/3}}\( \frac{ p_4^2{ -} \v p_2 {\cdot} \v p_4}{|p_4^2 {-} \v p_2 {\cdot} \v p_4|^{4/3}}+ \frac{ p_2^2{ -} \v p_2 {\cdot} \v p_4}{|p_2^2 {-} \v p_2 {\cdot} \v p_4|^{4/3}}\)~,
  \eea
  where to obtain the second form of the equality we made use of the relations in (\ref{A5id}). The $\text{Re }\mL_+$ in (\ref{Lplus2}) is quoted in the main text in (\ref{plus}) and the $\text{Re } \mL_-$ in (\ref{Lminus2}) appears in the main text in (\ref{minus}). 

\subsection{Integrating over angles} \label{apA4}
In this section we demonstrate the validity of (\ref{sgnA}). Specifically, we need to perform the angular integrals of $\mL_{\pm}$,
\be \label{sgnAv2}
U_{\mL_{\pm}} \equiv \int d\Omega_1 d\Omega_2 d\Omega_3 d\Omega_4\, \text{Re }\mL_{\pm} \delta(\v p_{12;34})~.
 \ee
 We start with the $\mL_-$ contribution. Since $\mL_-(1,2,3,4)$ only depends on the angle between $\v p_2$ and $\v p_4$, we may evaluate the integral over $\Omega_1$ and $\Omega_3$, 
 \be \label{A31}
 U_{\mL_{-}}= \int d\Omega_2 d\Omega_4\,  \text{Re }\mL_- (1,2,3,4)\Delta(|\v p_2{-}\v p_4|, p_1, p_3)^{-1}, 
 \ee
where we defined
 \be \label{B19}
\Delta(k, k_a, k_b)^{-1} \equiv \int d\Omega_a d\Omega_b \,\delta^3( \v k - \v k_a - \v k_b)  = 2\pi (k k_a k_b)^{-1}
\ee
as the integral of the three dimensional delta function.

 Since $ \text{Re }\mL_- (1,2,3,4)$ and $\Delta(|\v p_2{-}\v p_4|, p_1, p_3)$ only depend on the relative angle between $ \v p_2$ and $\v p_4$, which we denote by $\theta$ ($\v p_2{\cdot} \v p_4 = p_2 p_4 \cos \theta$), we may trivially perform the other angular integrals in (\ref{A31}). Thus we are left with, in $d=3$, using the explicit form of $\text{Re } \mL_-$ (\ref{Lminus2}), 
\be \label{B20}
\sgn\, U_{\mL_-} =\sgn\,   \int_{-1}^1 d (\cos \theta)\frac{-\lam}{|p_2{-}\v p_4|^{5/3}}\( \frac{ p_4^2{ -} \v p_2 {\cdot} \v p_4}{|p_4^2 {-} \v p_2 {\cdot} \v p_4|^{4/3}}+ \frac{ p_2^2{ -} \v p_2 {\cdot} \v p_4}{|p_2^2 {-} \v p_2 {\cdot} \v p_4|^{4/3}}\) =  -\sgn \lam~, 
\ee
where we established the last equality by defining $x\equiv p_2/p_4$ and $t \equiv \cos \theta$, so that $(\v p _2 {-}\v p_4)^2 = p_2 p_4 (x {+} x^{-1} {-} 2t)$, and noting that the following integral is  positive,
\be
\int_{-1}^1 dt \frac{1}{|x{+} x^{-1}{-} 2t|^{5/6}} \(\frac{x^{-1}{-}t}{|x^{-1}{-}t|^{4/3}}+\frac{x{-}t}{|x{-}t|^{4/3}}\) \geq 0
\ee
for all positive $x$, as  can easily be checked by numerically evaluating the integral.

Now let us look at the contribution of the $\mL_+$ term, 
 \be \label{U1234Lp}
U_{\mL_+}= \int d\Omega_1 d\Omega_2\,  \text{Re }\mL_+(1,2,3,4)\Delta(|\v p_1{+}\v p_2|, p_3, p_4)^{-1}, 
 \ee
 where we performed the integral over angles $\Omega_3$ and $\Omega_4$, since $ \text{Re }\mL_+(1,2,3,4)$ only depends on the angle between $\v p_1$ and $\v p_2$.
 Similar to the $\mL_-$ case, the piece   of $\Delta$ that depends on the angle between $\v p_1$ and $\v p_2$ is
 \be
 \Delta(|\v p_1{+}\v p_2|, p_3, p_4)^{-1}\sim |\v p_1 {+}\v p_2|^{-1}~~.
 \ee
The explicit form $\text{Re }\mL_+$ in three dimensions was given in (\ref{Lplus2}). The expression is given in terms of $\t p_{\pm}$ defined in (\ref{tpm}), which we may write in dimensionless variables $x=p_1/p_2$ and $t$ as, 
\be
\t p_{\pm} = p_1 p_2 \( \sqrt{x + x^{-1} + 2t}\pm \sqrt{x + x^{-1} - 2t}\)~, \ \ \ \ p_+^2 = p_1 p_2 (x + x^{-1} +2 t)~.
\ee
Making use of (\ref{B19}) and the following identity, valid for all positive $x$, 
\small
\be
\!\!\!\!\!\!\int_{-1}^1 dt \frac{1}{x{+}x^{-1}{+}2t} \[\!\Big( \sqrt{x{+}x^{-1}{+}2t}+\sqrt{x{+}x^{-1}{-}2t}\Big)^{-1/3}+\frac{ \sqrt{x{+}x^{-1}{+}2t}-\sqrt{x{+}x^{-1}{-}2t}}{| \sqrt{x{+}x^{-1}{+}2t}-\sqrt{x{+}x^{-1}{-}2t}|^{4/3}}\] \geq 0
\ee
\normalsize
establishes that $\sgn\, U_{\mL_+} = - \sgn \lam$.

\section{Kinetic equation at finite  $N$} \label{apB}
In this appendix we find the next-to-leading order  kinetic equation for a generalization of the nonlinear Schr\"odnger equation having $N$ fields: $\Psi_p^i$, where the index ranges from $i$ ranges from $1$ to $N$. 

These fields $\Psi_p^i$ are grouped into a vector $\vec \Psi_p$. The Hamiltonian is $O(N)$ invariant under rotations among the fields and is given by, 
\be \label{HN}
H=  \sum_p \o_p \, {\vec \Psi}^{\, *}_p{\cdot} \vec \Psi_p + \frac{\lam}{N}\sum_{p_1, \ldots, p_4}({\vec \Psi}^{\, *}_{p_1}{\cdot} \vec \Psi_{p_3})({\vec \Psi}^{\, *}_{p_2} {\cdot} \vec \Psi_{p_4})
\ee
in momentum space, or equivalently by (\ref{largeNH}) in position space. 
The relation between the occupation numbers and the fourth moment, generalizing (\ref{eomKE}), is 
\be \label{KE1B2}
\frac{\d n_1}{\d t} = -\frac{4 \lam}{N^2} \, \int d\v p_2 d\v p_3 d\v p_4\,\delta(\v p_{12;34})\Im \la  {\vec \Psi}_{p_1}{\cdot }  {\vec\Psi}^*_{p_3}\,   {\vec\Psi}_{p_2}{\cdot}  {\vec \Psi}^*_{p_4}\ra~,
\ee
where $n_k \equiv \frac{1}{N}\la  {\vec\Psi_k}{\cdot}  {\vec \Psi_k}\ra$ and both $n_1$ on the left and the correlation function on the right are evaluated at the same time. 

\subsubsection*{Leading order kinetic equation}
Let us first look at the correlation function at leading order in $\lam$, 
\be \label{Tree4pt}
\la \Psi_{1}^i \Psi_{2}^j \Psi_{3}^{*\, i} \Psi_{4}^{*\, j}\ra(t) = \frac{2 \lam}{N} n_1 n_2 n_3 n_4 \Big(\frac{1}{n_1} {+}\frac{1}{n_2}{-}\frac{1}{n_3}{-}\frac{1}{n_4}\Big)\frac{1}{\o_{p_3p_4;p_1p_2} {+}i\eps} \(1+ \delta_{i j}\)~,
\ee
where we are using shorthand for the subscripts, $\Psi_1^i\equiv \Psi_{p_1}^i $ and $n_1 \equiv n_{p_1} $. In the case that $i=j$, the new factor of $1+ \delta_{i j}$ equals $2$, and this is the same as the four-point function for the $N=1$ case of one field \cite{RS1}. For $i\neq j$, the result is smaller by a combinatorial factor of two, as a result of having to Wick contract fields with the same index. Inserting (\ref{Tree4pt}) into (\ref{KE1B2}) and summing over $i, j$ gives, 
\be \label{treeN}
\frac{\d n_1}{\d t} = \frac{8 \pi \lam^2}{N} \int d\vec p_2 d\vec p_3 d\vec p_4\,  n_1 n_2 n_3 n_4 \Big(\frac{1}{n_1} {+}\frac{1}{n_2}{-}\frac{1}{n_3}{-}\frac{1}{n_4}\Big)\Big( 1+ \frac{1}{N}\Big) \delta(\o_{p_1p_2; p_3 p_4})\delta(\v p_{12;34}) ~.
\ee
For $N=1$, this reproduces the earlier results (\ref{KE1}). 

\subsubsection*{Next-to-leading order kinetic equation}
Consider now the one-loop (next order in $\lam$) term in the kinetic equation. In the case of $N=1$, the diagram in Fig.~\ref{tree}(b) had a combinatorial factor of $8$. If $i\neq j$, this becomes $4$ (and is $1/N$ suppressed). Likewise (for $i\neq j$), Fig.~\ref{tree}(c) had a combinatorial factor of $16$, which now becomes $4$, and Fig.~\ref{tree}(d) had a combinatorial factor of $16$, which now becomes $0$. Fig.~\ref{tree}(b) gives $\mL_+$ and  Fig.~\ref{tree}(c) and Fig.~\ref{tree}(d) give $\mL_-$. Therefore, while for $N=1$ we had in the kinetic equation
\be
4(1 + 2\mL_+ + 8 \mL_-)
\ee
we now have the contribution (from $i\neq j$), 
\be \label{19}
2N (N{-}1)\( \frac{1}{N}+ \frac{2}{N^2} \mL_+ +\frac{2}{N}\mL_-\)~,
\ee
where we accounted for the fact that now each vertex comes with a factor of $1/N$, and there are $N(N{-}1)$ contributions when summing over all $i \neq j$. 
There is also the following contribution from  $i=j$:
\be \label{110}
4 N\(\frac{1}{N}+ \frac{2}{N^2} \mL_+ +\frac{8}{N^2} \mL_- \)  + 2\frac{N{-}1}{N} \mL_-~,
\ee
where the first term is the same as the $N=1$ answer (but accounting for the $1/N$ factor at each vertex and multiplying by an overall $N$ from the sum over all $i$), and the second term comes from cases in which the index $k$ inside the loop with arrows in opposite directions does not equal index $i$ (where the prefactor has $1/N^2$ from the $1/N$ in each vertex, multiplied by an $N$ from summing all $i$, and an $N{-}1$ from summing over all $k\neq i$, and a factor of $2$ because both Fig.~\ref{tree}(c) and (d) contribute). 
Adding both contributions (\ref{19}) and (\ref{110}) we get that the kinetic equation is
\bea\nn 
\frac{\d n_1}{\d t} = \frac{8 \pi \lam^2}{N} \text{Re} \int d\vec p_2 d\vec p_3 d\vec p_4 \, n_1 n_2 n_3 n_4 \Big(\frac{1}{n_1} {+}\frac{1}{n_2}{-}\frac{1}{n_3}{-}\frac{1}{n_4}\Big)\delta(\o_{p_1p_2; p_3 p_4})\delta(\v p_{12;34})\\ \label{KEnlo}
\Big[ \Big(1+ \frac{1}{N}\Big)\! \Big(1+ \frac{2 \mL_+}{N}\Big) +  \Big(2 -\frac{1}{N}+ \frac{15}{N^2}\Big)\mL_-  \Big]~,
\eea
where $\mL_{\pm}$ are given by (\ref{A2}). 
For $N=1$ this reduces to (\ref{12}), while for large $N$ this becomes, 
\be \label{112}
\frac{\d n_1}{\d t} = \frac{8 \pi \lam^2}{N} \text{Re} \int d\vec p_2 d\vec p_3 d\vec p_4\,  n_1 n_2 n_3 n_4 \Big(\frac{1}{n_1} {+}\frac{1}{n_2}{-}\frac{1}{n_3}{-}\frac{1}{n_4}\Big)\(1+2 \mL_- \)\delta(\o_{p_1p_2; p_3 p_4})\delta(\v p_{12;34})~.
\ee
The large $N$ kinetic equation (valid at large leading  nontrivial order in $1/N$, and at all orders in $\lam$), was given in the main body,  (\ref{largeNKE}). Expanding  (\ref{largeNKE}) to order $\lam$ reproduces the one loop result (\ref{112}).

 \section{Scale invariant solutions} \label{apD}
 In this appendix we review some properties of self-similar time-dependent solutions, which appear in Sec.~\ref{selfMain}. 
 
 One expects that either at very weak coupling or at very strong coupling, and far away from the forcing and dissipation scales, the kinetic equation has solutions that are scale invariant, taking the form, 
\be \label{51}
n_k = t^a f(k t^b)~.
\ee
The exponents $a$ and $b$ are not independent. If one has a flux with constant particle number (freely decaying turbulence) one finds the relation,
\be \label{52}
N = \int d^d k n_k = t^{a-db}\int d^d k  f(k)~, \ \ \ \ \Rightarrow  a- bd =0~,
\ee
where we did a change of variables $k \rightarrow k/t^b$.
If, instead, the particle number grows linearly with time (forced turbulence) we have the relation, 
\be \label{53}
N\sim t~, \ \ \ \ \ \ \ \Rightarrow  a- bd =1~.
\ee
To establish a second relation between $a$ and $b$ we must look at the kinetic equation. We do this first in the weak coupling limit (which corresponds to the UV for the particle number cascade in the nonlinear Schr\"odinger equation), and then the strong coupling limit of the large $N$ kinetic equation (which corresponds to the IR for the particle number cascade in the nonlinear Schr\"odinger equation).

\subsection*{Weak nonlinearity}
Let us look at the standard (weak nonlinearity) kinetic equation. We take a general dispersion relation $\o_k \sim k^{\al}$ and interaction $\lam_k \sim k^{\beta}$.  The scaling for the kinetic equation (e.g. (\ref{27}) with $\Lambda_{1234} \sim k^{\beta}$) is of the form, 
\be \label{56}
\frac{\d n_k}{\d t} \sim \frac{k^{2d} n_k^{3} \lam_k^2}{\o_k}~.
\ee
In order for the $t$ dependence to cancel (and noting that $k\sim 1/t^{b}$), we must have 
$
t^{ a-1} \sim t^{3a} (t^{-b})^{2(d+ \beta) - \al}
$.
Thus, 
\be \label{58}
2a+1 = b(2d + 2\beta - \al)~.
\ee
Combining with either (\ref{52}) or (\ref{53}) we therefore have the solutions (for the particle number cascades), 
\bea \label{59}
a &=& \frac{d}{2\beta - \al}~, \ \ \ \ \ \ \  \ b = \frac{1}{2\beta - \al}~, \ \ \ \ \ \text{freely decaying } \\ \label{510}
a &=&1+ \frac{3d}{2\beta - \al}~, \ \ \ b = \frac{3}{2\beta - \al}~, \ \ \ \ \ \text{forced}~.
\eea
Notice that a consistency check is to take $f(\xi) = \xi^{-a/b}$. In this case $n_k = k^{-a/b}$, and so $a/b$ should match the KZ scaling we know. Indeed, in the forced case we get $\frac{a}{b}= d + \frac{2}{3}\beta - \frac{\al}{3}$, matching the KZ scaling. In the freely decaying case we instead have $a/b = d$. 
We may now specialize to our case of the nonlinear Schr\"odinger equation, which has $\al = 2$ and $\beta =0$. This gives, 
\be
n\sim t^{- d/2} f\(\frac{k}{\sqrt{t}}\)~,  \ \ \text{freely decaying}~, \ \ \ \ \ \ \ \& \ \ \ \ \ \ \ \  n\sim t^{1-\frac{3d}{2}} f\(\frac{k}{t^{\frac{3}{2}}}\)~, \ \ \ \ \ \text{forced}~.
\ee

Suppose now that we look for power law solutions of the kinetic equation, $
f(x) \sim k^{-\g} $.
We can establish $\g$ from (\ref{56}). Since $k$ scales like $x$, and since all the $t$ dependence has already been canceled, we get
\be
x^{-\g} \sim x^{2d - 3\g + 2\beta -\al}~, \ \ \ \ \ \ \Rightarrow \g = d + \beta -\frac{\al}{2}~.
\ee
This corresponds to 
\be \label{517}
n_k \sim t^{a - b(d+ \beta -\frac{ \al}{2})} k^{- (d+ \beta - \frac{\al}{2})}\sim t^{- 1/2}k^{- (d+ \beta - \frac{\al}{2})}
\ee
for both the freely decaying and forced case, where we made use of  (\ref{59}) and (\ref{510}) to get the second equality. In the case of the nonlinear Schr\"odinger equation this becomes $n\sim t^{-1/2} k^{-d+1}$.

\subsection*{Strong nonlinearity}
If one takes the effective interaction at strong nonlinearity to be of the form (\ref{Lamstrong}), then schematically it takes the form, 
\be
|\Lambda_{1234}|^2 \approx \frac{\lam^2}{|\mL_-|^2}\sim \frac{\o_k^2}{(n_k k^d)^2}~, \ \ \ \ \Rightarrow  \frac{\d n_k}{\d t} \sim n_k \o_k~,
\ee
where in the last equality for $\Lambda_{1234}$ we wrote the schematic form, making use of the form of $\mL_-$ (\ref{A2}). 
Plugging in a scale invariant solution (\ref{51}) into the above schematic kinetic equation, in order for the time dependence to cancel on both sides, 
$
t^{a-1} \sim t^{a-b\al}$, which means $b = \frac{1}{\al}$. 
Combining with either (\ref{52}) or (\ref{53}) we therefore have the solutions, 
\bea
a &=& \frac{d}{\al}~, \ \ \ \ \ \ \   \ b = \frac{1}{ \al}~, \ \ \ \ \ \text{freely decaying } \\
a &=&1+ \frac{d}{\al}~, \ \ \ b = \frac{1}{ \al}~, \ \ \ \ \ \text{forced}~.
\eea
Specializing to our case of the nonlinear Schr\"odinger equation, which has $\al = 2$ and $\beta =0$, 
\be
n\sim t^{d/2} f(k\sqrt{t})~, \ \ \ \ \ \text{freely decaying}~, \ \ \ \ \ \ \ \ \ \ \ 
n\sim t^{1+\frac{d}{2}}  f(k\sqrt{t})~, \ \ \ \ \ \text{forced}~.
\ee
The stationary solutions are therefore of the form $n_k \sim k^{-a/b}$,
\be
n_k \sim k^{-d}~, \ \ \ \text{freely decaying }~, \ \ \ \ \ \ \ \ \ \ \ n_k\sim k^{-d- \al} = k^{-d -2}~, \ \ \ \text{forced}~,
\ee
where in the last equality we specialized to the nonlinear Schr\"odinger equation. \\

However, as noted below (\ref{ss2}) $n_k\sim k^{-d -2}$ causes $\mL_-$ to diverge in the IR, so this solution is not self-consistent. We instead do the following: we find $n_k$ at strong coupling by assuming $n_k \sim k^{-\g}$ (for small $k$) with $\g>d$, implying an IR divergence in $\mL_-$. Using the effective interaction (\ref{BigLam}) with  scaling of $\beta=2$ gives that the stationary solution at strong coupling is just the  (weak coupling) Kolmogorov-Zakharov solution, $n_k \sim k^{-d -2/3 \beta +\al/3}$, with $\beta = 2$ and $\al=2$. This is $n_k \sim k^{- d -2/3}$, as quoted in the main body of the text, (\ref{ss3}). Likewise, for the time-dependent solution (\ref{517}), inserting $\beta = \al = 2$ gives, $n_k \sim t^{-1/2} k^{-d -1}$ (\ref{ss1}). 

\bibliographystyle{utphys}
%\bibliography{LargeNGPTurbulenceBib}

\end{document}